\documentclass[10pt, aps, prd, superscriptaddress, nofootinbib, showpacs, twocolumn,longbibliography]{revtex4-2}
\usepackage[english]{babel}
\usepackage{amsmath, amssymb, amsfonts, amsthm, mathrsfs}
\usepackage[all]{xy}
\usepackage{bm}
\usepackage{stmaryrd}
\usepackage{graphicx}
\usepackage{hyperref}
\usepackage{color}

\hypersetup{breaklinks=true}

\usepackage{pgfplots, grffile, tikz}
\pgfplotsset{compat=newest}
\usetikzlibrary{plotmarks, arrows.meta}
\usepgfplotslibrary{patchplots}

\numberwithin{equation}{section}

\DeclareMathOperator{\Tr}{Tr}

\newcommand{\calE}{\mathcal E}

\newcommand{\calR}{\mathcal R}

\newcommand{\bbG}{\mathbb G}

\newcommand{\bbK}{\mathbb K}

\newcommand{\lb}{\llbracket}
\newcommand{\rb}{\rrbracket}

\newcommand{\be}{\begin{equation}}
\newcommand{\ee}{\end{equation}}
\newcommand{\bl}{\begin{align}}
\newcommand{\el}{\end{align}}

\newcommand{\al}{\alpha}

\newcommand{\p}{\partial}

\newcommand{\lp}{\left(}
\newcommand{\rp}{\right)}
\newcommand{\lc}{\left\{}
\newcommand{\rc}{\right\}}

\newcommand{\h}{\hat}
\renewcommand{\O}{\varOmega}
\newcommand{\nb}{\nabla}

\begin{document}
% ======================================

\title{Commutator technique for the heat kernel of minimal higher derivative operators}

\author{A.\,O. Barvinsky}
\email{barvin@td.lpi.ru}
\affiliation{Theory Department, Lebedev Physics Institute, Leninsky Prospect 53, Moscow 119991, Russia}

\author{A.\,V. Kurov}
\email{kurov@td.lpi.ru}
\affiliation{Theory Department, Lebedev Physics Institute, Leninsky Prospect 53, Moscow 119991, Russia}

\author{W. Wachowski}
\email{vladvakh@gmail.com}
\affiliation{Theory Department, Lebedev Physics Institute, Leninsky Prospect 53, Moscow 119991, Russia}

\begin{abstract}
We suggest a new technique of the asymptotic heat kernel expansion for minimal higher derivative operators of a generic $2M$-th order, $F(\nabla)=(-\Box)^M+\cdots$, in the background field formalism of gauge theories and quantum gravity. This technique represents the conversion of the recently suggested Fourier integral method of generalized exponential functions [Phys. Rev. D105, 065013 (2022), arXiv:2112.03062] into the commutator algebra of special differential operators, which allows one to express expansion coefficients for $F(\nabla)$ in terms of the Schwinger-DeWitt coefficients of a minimal second order operator $H(\nabla)$. This procedure is based on special functorial properties of the formalism including the Mellin-Barnes representation of the complex operator power $H^M(\nabla)$ and naturally leads to the origin of generalized exponential functions without directly appealing to the Fourier integral method. The algorithm is essentially more straightforward than the Fourier method and consists of three steps ready for a computer codification by symbolic manipulation programs. They begin with the decomposition of the operator into a power of some minimal second order operator $H(\nabla)$ and its lower derivative ``perturbation part'' $W(\nabla)$, $F(\nabla)=H^M(\nabla)+W(\nabla)$, followed by considering their multiple nested commutators. The second step is the construction of special local differential operators---the perturbation theory in powers of the lower derivative part $W(\nabla)$. The final step is the so-called procedure of their {\em syngification} consisting in a special modification of the covariant derivative monomials in these operators by Synge world function $\sigma(x,x')$ with their subsequent action on the HaMiDeW coefficients of $H(\nabla)$.
\end{abstract}

\maketitle

%%%%%%%%%%%%%%%%%%%%%%%%%%%%%%%%%%%%%
\section{Introduction}
%%%%%%%%%%%%%%%%%%%%%%%%%%%%%%%%%%%%%
Local curvature and gradient expansion in quantum gravity theory is largely determined by the properties of the heat kernel for various differential operators -- inverse propagators of quantum fields on curved space background. Usually it is enough to have the functional trace of the heat kernel to determine the one-loop effective action of the theory, but the construction of Green's functions, multi-loop contributions to the effective action within the background field formalism, etc., requires the knowledge of the two-point (off-diagonal) heat kernel governed by the Schwinger-DeWitt technique \cite{Schwinger,DeWitt1965,Barvinsky1985,Scholarpedia}. Moreover, the basic tool of this technique---Schwinger-DeWitt expansion of the off-diagonal heat kernel for {\em minimal second order} operators, is in fact needed for obtaining the recursive algorithm for the coefficients of its diagonal elements participating in its functional trace \cite{DeWitt1965}. At the same time, there is a long list of field theory and quantum gravity models with nonminimal and higher derivative wave operators \cite{Stelle,fradkin1982,Avramidy1985,Codello:2006in,Percacci:2009fh,Salvio:2014soa, Alvarez-Gaume:2015rwa,Anselmi:2018ibi, Rachwal:2021bgb,Tseytlin:2022flu,GrassoKuzenko2023,Donoghue:2023yjt,Holdom:2023usn,Buccio:2024hys}.

The extension of Schwinger-DeWitt technique to such models was recently built in \cite{Wach3}. This work generalized the method of \cite{Schwinger,DeWitt1965,Barvinsky1985,Fulling,Gusynin1989, Gusynin1990, Gusynin1991, GusyninGorbar, GusyninGorbarRomankov} to a wide class of operators formally described by mathematical heat kernel theory of Gilkey-Seeley \cite{Seeley, Gilkey1975, Gilkey1979, Vassil03,Fursaev:2011zz}. In the case of minimal operators $\hat F(\nabla)$ of any order it suggested explicit workable algorithms of the covariant expansion for their two-point (off-diagonal) heat kernels,
\begin{equation}
\hat K_F(\tau|x,x') = \exp\left(-\tau\hat F(\nabla_x)\right) \delta(x,x')\,g^{-1/2}(x'), \label{HK0}
\end{equation}
while for generic operators it reduced the procedure of this expansion to the calculation of certain matrix-valued Fourier integrals. Both results were based in \cite{Wach3} on the covariant version of the Fourier transform, adjusted to curved space applications and incorporating a fully covariant method of Synge world function.

The main achievement of the work \cite{Wach3} was that it has overstepped the limitations of the Schwinger-DeWitt method \cite{Schwinger,DeWitt1965,Barvinsky1985} which restricts the applicability of the well-known expansion of the heat kernel
\begin{multline} \label{HeatKernel0}
\hat K_H(\tau|x,x') = \frac{\Delta^{1/2}(x,x')}{(4\pi\tau)^{d/2}}
\sum\limits_{m=0}^\infty \tau^{m}\\
\times \exp\left(-\frac{\sigma(x,x')}{2\tau}\right)\,\hat a_m(H|x,x'),
\end{multline}
to the case of {\em minimal second-order} differential operators of the form
\begin{equation} \label{minimal2}
\hat H(\nabla) = -\Box\,\hat 1 + \hat P,
\end{equation}
whose derivatives form a covariant Laplacian $\Box = g^{ab}\nabla_a\nabla_b$ defined in a curved $d$-dimensional spacetime with the Riemannian metric $g_{ab}(x)$ and $\hat P=\hat P(x)$ forms the potential term. Here the covariant derivatives $\nabla_a$ (and the operator as a whole) are acting on a generic set of fields $\varphi(x)=\varphi^A(x)$ with spin-tensor labels $A$ of arbitrary nature, and the hat denotes matrices in the vector space of $\varphi^A$, in particular $\hat 1=\delta^A_B$ denoting the corresponding unit matrix. Only in this particular case there exists the Schwinger--DeWitt ansatz (\ref{HeatKernel0}) for its heat kernel (\ref{HK0}) dedensitized by $g^{-1/2}(x')$ in its second argument.\footnote{We assume that the delta-function defined by the relation $\int d^dx\,\delta(x,x')\varphi(x')=\varphi(x)$ is the density in its second argument and work in the Euclidean space version of the theory related to the physical Lorentzian signature spacetime by Wick rotation, so that $-\Box$ is positive-definite under appropriate boundary conditions. The expression (\ref{minimal2}) also embraces the case of a linear in derivatives term because it can always be absorbed into the $\Box$-term by the redefinition of connection in $\nabla_a$.}

Other ingredients of the expansion (\ref{HK0}) are the Synge world function $\sigma(x,x')$---one half of the square of the geodetic distance between the points $x$ and $x'$, the dedensitized Pauli-Van Vleck-Morette determinant $\Delta(x,x')=\det(-\partial_a\partial_{b'}\sigma)/g^{1/2}(x)\,g^{1/2}(x')$ and well-known HaMiDeW or Schwinger--DeWitt coefficients $\hat a_m(x,x')$. These coefficients satisfy recurrent differential equations which can be successively solved for $\hat a_m(x,x')$ in the form of covariant Taylor series in powers of $\sigma^{a'} = \nabla^{a'}\!\sigma$---the vector tangential at the point $x'$ to the geodetic connecting $x$ and $x'$. The coefficients of this Taylor expansion are local functions of spacetime metric, its curvature and background fields, and thus provide all the goals of perturbative UV renormalization of local field models and their effective field theory expansion.

Covariant Fourier transform of \cite{Wach3} suggests the generalization of (\ref{HK0}) to the case of generic higher derivative operator
%\begin{align} \label{operator}
$\hat F(\nabla) = \sum_{k=0}^{N} \hat F_k^{a_1\ldots a_k}(x)\nabla_{a_1}\cdots\nabla_{a_k}$.
%\end{align}
For the case of the {\em minimal differential operator of order $2M$},
\begin{align}
&\hat F(\nabla) = \hat 1 (-\Box)^M + \hat W(\nabla), \label{MinOpDef}
%\\
%&\hat P(\nabla) = \sum\limits_{k=0}^{2M-1} \hat P_k^{a_1\ldots a_k}(x)\nabla_{a_1}\cdots\nabla_{a_k},
\end{align}
which has the leading term proportional to a power of the covariant Laplacian $\Box$ and a generic part $\hat W(\nabla)$ of lower order $2M-1$ in derivatives, this generalization turns out to be most constructive (see Eq.(2.21) of \cite{Wach3}).

Mainly it consists in the replacement of integer powers of the proper time by fractional powers $\tau^{m/M}$ and the single summation gets replaced by the double sum $\sum_{m,k}$ with a finite range of integer $k$ depending on $m$. This sum runs over a special set of two-point functions $\hat b_{m,k}(x,x')$ weighted by what was called in \cite{Wach2} {\em generalized exponential functions} $\calE_{M,d/2+M k-m}(z)$ of the argument $z=-\sigma/2\tau^{1/M}$. The set of these functions, parameterized by two indices and defined by the series
\begin{equation}\label{calE}
\calE_{\nu,\alpha}(z) = \frac{1}{\nu} \sum\limits_{k=0}^\infty
\frac{\Gamma\left(\frac{\alpha+k}{\nu}\right)}{\Gamma(\alpha+k)}\frac{z^k}{k!},
\end{equation}
replaces one overall single exponential function $\exp(-\sigma/2\tau)$ in (\ref{HeatKernel0}). The properties of $\calE_{\nu,\alpha}(z)$ were studied in much detail in \cite{Wach2}.

Important distinction from the conventional Schwinger-DeWitt expansion (\ref{HK0}) is that fractional powers of the proper time for the operators (\ref{MinOpDef}) include negative values and extend to minus infinity, $-\infty<m<\infty$ \cite{Wach3}. In the coincidence limit $x'=x$, however, the coefficients of the negative powers vanish in full accordance with the Gilkey-Seeley theory of the heat kernel trace \cite{Seeley, Gilkey1975, Gilkey1979}.

By and large the generalized coefficients $\hat b_{m,k}(x,x')$ of \cite{Wach3} just like original HaMiDew coefficients $\hat a_m(x,x')$ have a growing with $m$ and $k$ dimensionality $dim\,\hat b_{m,k}\geq\max\{2m, k/2\}$, so that any given order of curvature and gradient expansion can be constructed in terms of finite set of $\hat b_{m,k}(x,x')$ and this expansion turns out to be efficient for the purpose of effective field theory below some energy cutoff. However, the Fourier integral algorithms of their construction (see Eqs.(2.23),(2.6),(3.22)-(3.24), etc. of \cite{Wach3}) turned out to be very involved and hard for implementation, in particular, because of the necessity to consider a recursive procedure (Eq.(3.27) of \cite{Wach3}) for an auxiliary initial value problem.

Thus, the goal of this paper will be to show that the difficulties of the above procedure can be circumvented by converting the Fourier integral method of \cite{Wach3} into a special commutator algebra of differential operators. This algebra allows one to express heat kernel expansion coefficients of the minimal higher derivative operator (\ref{MinOpDef}) in terms of the Schwinger-DeWitt coefficients $\hat a_m(x,x')$ of the second order one (\ref{minimal2})---the procedure based on special functorial properties of the formalism. This includes Mellin-Barnes representation of the complex power of the operator \cite{Wach3}, which naturally leads to the origin of generalized exponential functions (\ref{calE}) without directly appealing to the Fourier integral method.

The whole algorithm is essentially more efficient than the Fourier method of \cite{Wach3} and consists of three straightforward steps ready for a simple computer codification by symbolic manipulation programs like xAct package of Wolfram {\em Mathematica}. They begin with splitting the operator into a power of some minimal second order one, $H^M(\nabla)$---not necessarily $(-\Box)^M$, and its lower derivative ``perturbation part'' followed by considering multiple nested commutators of these two parts. The second step is the construction of special local differential operators---the perturbation theory in powers of the lower derivative part of the operator. The final step is the so-called procedure of their {\em syngification} associated with the name of John Synge, the author of the world function $\sigma(x,x')$. This step consists in a special modification of the covariant derivative monomials of these operators by $\sigma(x,x')$ with their subsequent action on the HaMiDeW coefficients of $H^M(\nabla)$.

The paper is organized as follows. Sect.\ref{SummarySect} contains the summary of results, Sect.\ref{PertTheory} gives their derivation, while Sects.IV and V present examples of applications to a simple 4-th order operator and the 6-th order operator arising in renormalizable Ho\v rava gravity theory along with the verification of low-order Gilkey-Seeley coefficients for the heat kernel trace of a generic 4-th order operator. Appendices A, B and C contain the derivation of the functorial property for a complex power of the minimal second order operator, special syngification identity and tables of simplest syngified monomials of covariant derivatives.

%%%%%%%%%%%%%%%%%%%%%%%%%%%%%%%%%%%%%
\section{Summary of results \label{SummarySect}}
%%%%%%%%%%%%%%%%%%%%%%%%%%%%%%%%%%%%

Each minimal operator (\ref{MinOpDef}) of order $2M$ can obviously be decomposed as the sum of two terms -- $M$-th power of some minimal second order operator $\hat H(\nabla)$ of the form (\ref{minimal2}) and the corresponding complemetary term $\hat W(\nabla)$ which is the operator of the lower order $2M-1$,
\begin{eqnarray}
&&\hat F(\nabla) = \hat H^M(\nabla) + \hat W(\nabla), \label{PowerDecomposition} \\
&&\hat W(\nabla) = \sum\limits_{k=0}^{2M-1} \hat W_k^{a_1\ldots a_k}(x)\nabla_{a_1}\cdots\nabla_{a_k}. \label{Perturbation}
\end{eqnarray}
The idea of this decomposition is that $\hat W(\nabla)$ will be treated by perturbation theory, whereas $\hat H^M(\nabla)$ will comprise its leading order. This decomposition is of course not unique and depends on the choice of the potential term $\hat P$ in (\ref{minimal2}). This choice can be used to simplify calculations in concrete applications and will not be specified otherwise until the examples of calculations in Sects.IV and V (where $\hat H(\nabla)$ will be simply identified with $-\Box$).

The virtue of the decomposition (\ref{PowerDecomposition}) is that it allows one to reduce the heat kernel expansion of \cite{Wach3}, based on a complicated generalized Fourier transform, to a rather straightforward commutator algebra of the operators $\hat H^M(\nabla)$ and  $\hat W(\nabla)$ and reexpress the result in terms of the well-known Schwinger-DeWitt coefficients $\hat a_l(H|x,x')$ of the minimal second order operator $\hat H(\nabla)$. These coefficients turn out to be acted upon by certain explicitly calculable local differential operators originating from this commutator algebra.

Thus, the main result of the paper can be formulated as follows. The heat kernel of a higher-order minimal operator \eqref{PowerDecomposition} reads as a double series expansion in fractional powers of the proper time $\tau^{\pm m/M}$, $m=0,1,\cdots \infty$, and generalized exponential functions (\ref{calE}), $\calE_{M,\alpha}(z)$, of the argument $z=-\sigma(x,x')/2\tau^{1/M}$ summed over various values of the order parameter $\alpha$,
\begin{widetext}
\begin{eqnarray} \label{GenExpansion}
\hat K_F(\tau|x,x')\,&=&\, \frac{1}{(4\pi\tau^{1/M})^{d/2}}\sum\limits_{m=0}^\infty\; \tau^\frac{m}{M}\;
\sum\limits_{k=0}^\infty\;
\calE_{M,\frac{d}{2}+M k-m}\Big(\!-\frac{\sigma(x,x')}{2\,\tau^{1/M}}\Big)\, \hat a_{m,k}(F|x,x')\nonumber\\
&+&\,\frac{1}{(4\pi\tau^{1/M})^{d/2}}\sum\limits_{m=1}^{\infty}\; \tau^{-\frac{m}{M}}\!\!\!
\sum_{\;k\ge\frac{m}{M-1}}^\infty\!\!
\calE_{M,\frac{d}{2}+M k+m}\Big(\!-\frac{\sigma(x,x')}{2\,\tau^{1/M}}\Big)\, \hat a_{-m,k}(F|x,x').
\end{eqnarray}
The double-indexed coefficients of this expansion $\hat a_{m,k}(F|x,x')$ for both positive and negative $m$ are given by the expression
\begin{eqnarray}\label{amk}
&&\hat a_{m,k}(F|x,x')=\sum\limits_{l\ge L_{m,k}}^{m+(M-1)k}
\left\{\hat U_k(\nabla_x)\right\}_{M k+l-m} \Delta^{1/2}(x,x')\,\hat a_l(H|x,x'),\\
&&L_{m,k} = \max\big\{0, m-M k\big\},\quad -\infty<m<\infty,
\end{eqnarray}
\end{widetext}
in terms of the Schwinger-DeWitt coefficients $\hat a_l(H|x,x')$ of the operator $\hat H(\nabla)$ weighted by the Pauli-Van-Vleck factor and acted upon (on the argument $x$) by special differential operators $\{\hat U_k(\nabla_x)\}_{M k+l-m}$ mentioned above.

The construction of these operators consists of three steps. First, one introduces the multiple nested commutator of $k$-th order
\begin{equation}\label{Vk}
\hat V_k(\nabla) = \lb\hat W(\nabla), \hat H^M\rb_k,
\end{equation}
which is defined for any two operators $A$ and $B$ by the sequence of relations
\begin{align}\label{NestedComm}
\begin{aligned}
\big\lb\, A, B \big\rb_0 &\equiv A,\\
\big\lb\, A, B \big\rb_k
&\equiv \big[\ldots\big[[A,\underbrace{B], B\big],\ldots, B}_{k}\big].
\quad k>0.
\end{aligned}
\end{align}
From the structure of operators $\hat H^M(\nabla)$ and $\hat W(\nabla)$ it follows that $\hat V_k(\nabla)$ is a differential operator of order $(2M-1)(k+1)$ and has the physical dimension $2M(k+1)$---the sum of the dimensionality of derivatives and the dimensionality of their coefficients. Therefore the {\em background dimensionality} of $\hat V_k(\nabla)$---the minimal dimensionality of its coefficients---is
\be\label{dimVk}
dim\,\hat V_k(\nabla)=2M(k+1)-(2M-1)(k+1)=k+1,
\ee
so that the expansion in powers of $\hat V_k(\nabla)$ with a growing $k$ is efficient for obtaining the local gradient expansion of the heat kernel and relevant physical objects discussed in Introduction.

The second step on the road to the operators $\{\hat U_k(\nabla)\}_{M k+l-m}$ is the construction of the sequence of local differential operators $\hat U_k(\nabla)$ defined by the relations
\be\label{U0}
\hat U_0(\nabla)=\hat 1
\ee
and
\begin{align}\label{EvOpExp}
&\hat U_k(\nabla) =\!\!\! \sum\limits_{\overset{\scriptstyle n,k_1,\cdots k_n}{n+|\bm{k}|=k}}
(-1)^n\frac{\hat V_{k_1}(\nabla)\cdots \hat V_{k_n}(\nabla)}{k_1!\cdots k_n!\, c(\bm{k})},\;k>0,\\
&c(\bm{k}) = (k_1\!+\!1)(k_1\!+k_2\!+\!2)\cdots(k_1\!+\cdots+\!k_n\!+n),\label{c(k)}
\end{align}
where summation runs over the indicated domain of integer indices $n,k_1,\cdots k_n$, $n+|\bm{k}|=k$, and we use standard multi-index notations: $\bm{k} = (k_1,\ldots,k_n)$, $|\bm{k}| = k_1 + \cdots + k_n$. Obviously $\hat U_k(\nabla)$ are differential operators of order $(2M-1)k$, physical dimension $2M k$ and background dimensionality $dim\,\hat U_k(\nabla)=\sum_{i=1}^n\,dim\,\hat V_{k_i}=n+|\bm{k}|=k$.

The transition $\hat U_k(\nabla)\to \{\hat U_k(\nabla)\}_{M k+l-m}$ is the final step of our construction. It consists in the application of the so-called {\em syngification} operation, $\hat F(\nabla)\to\{\hat F(\nabla)\}_n$, denoted by curly brackets with the subscript $n=M k+l-m$. This operation can be defined for any $N$-th order local differential operator $\hat F(\nabla) = \sum_{k=0}^{N} \hat F_k^{a_1\ldots a_k}(x)\nabla_{a_1}\cdots\nabla_{a_k}$ as
\begin{equation} \label{CurlBrDef}
\{\hat F\,\}_n = \sum\limits_{j=n}^N \hat F^{a_1...a_j} \left\{\nabla_{a_1}...\nabla_{a_j}\right\}_n
\end{equation}
where $\{\nabla_{a_1}...\nabla_{a_j}\}_n$ represents the sum of all possible monomials $\nabla_{a_1}...\nabla_{a_j}$ in which $n$ covariant derivatives $\nabla_a$ are replaced by relevant derivatives of the Synge world function $-\nabla_a\sigma(x,x')/2=-\sigma_a(x,x')/2$. Such a replacement should be performed without changing the ordering of the remaining $(j-n)$ derivatives and $n$ newly introduced factors. This definition can be formalized as the following expression
\begin{eqnarray}\label{synge}
&&\left\{\nabla_{a_1}...\nabla_{a_j}\right\}_n=\frac1{n!}\frac{\partial^n}{\partial k^n}\,
e^{\frac{k\sigma(x,x')}2}\,\nabla_{a_1}...\nabla_{a_j}\,e^{-\frac{k\sigma(x,x')}2}\,\Big|_{\,k=0}\nonumber\\
&&\qquad=\frac1{2^n n!}\frac{\partial^n}{\partial k^n}\,\big(\nabla_{a_1}\!\!-k\sigma_{a_1}\big)...
\big(\nabla_{a_j}\!\!-k\sigma_{a_j}\big)\,\Big|_{\,k=0}.
\end{eqnarray}
Obviously, $\left\{\nabla_{a_1}...\nabla_{a_j}\right\}_n=0$ for $n>j$.

The syngification of the operator $\hat U_k(\nabla)$ implies that the result $\{\hat U_k(\nabla)\}_n$ is again a local differential operator with {\em nonlocal} coefficients containing the two-point Synge vector function $\sigma_a(x,x')$ and its derivatives. It is of order $(2M-1)k - n$ in derivatives, has physical dimension $2M k - 2n$ and background dimensionality
\be\label{dimUkn}
dim\,\{\hat U_k\}_n=k-n.
\ee

The expansion \eqref{GenExpansion} is a direct analogue of the expansion previously obtained in \cite{Wach3} by the Fourier integral method --- see Eq.(2.17) of \cite{Wach3}. However, instead of a rather complicated and messy recurrent procedure for its coefficients (Eq.(2.23) of \cite{Wach3}), now we have a straightforward commutator algebra algorithm ready for a simple computer codification.

Just like in \cite{Wach3} the expansion \eqref{GenExpansion} contains a nontrivial series with arbitrarily large negative powers of the proper time $\tau$, which was explicitly disentangled in \eqref{GenExpansion}. As it was noted in \cite{Wach3}, this makes impossible for the higher-order derivative case to have a chain of recurrence relations similar to the one that works for minimal second order operators. However, as it will be shown in what follows, the coincidence limit $x=x'$ is contributed only by the coefficients $\hat a_{m,k}(F|x,x')$ for which $k\le 2m$ or
\begin{equation}\label{NonvanishingCoeff}
[\,\hat a_{m,k}(F)\,]=0, \quad k>2m,
\end{equation}
where the square brackets denote the coincidence limit, $[\,\hat a_{m,k}(F)\,] = \hat a_{m,k}(F|x,x)$ -- the notation which will be used throughout the paper.
So the series with negative powers of $\tau$ in \eqref{GenExpansion} vanishes in the coincidence limit. In fact, this means that every negative power of $\tau$, except the prefactor $1/(4\pi\tau^{1/M})^{d/2}$, appears in the combination with a positive power of $\sigma_a(x,x')$ vanishing in this limit, so that the expansion \eqref{GenExpansion} is a double asymptotic series in two variables $\tau^{1/M}\to 0$ and $\sigma^{a'}(x,x')/\tau^{1/M}\to 0$.\footnote{Indeed, if we expand the coefficients $\hat a_{m,k}$ into a covariant Taylor series, the off-diagonal heat kernel becomes a function of two variables -- dimensional $\tau$ and dimensionless $\sigma^{a'}/\tau^{1/M}$, and its asymptotic expansion runs in powers of both of them tending to zero. Thus, infinite accumulation of negative powers of $\tau^{1/M}$ is the result of the expansion in $\sigma^{a'}/\tau^{1/M}$ which vanishes at $x'=x$. Curiously, for minimal second order operators such terms do not arise at all, which apparently follows from the fact that the Schwinger-DeWitt series (\ref{HK0}) turns out to be a semiclassical expansion which remains homogeneous in the vicinity of $x'\to x$ \cite{Wach2}.}

Passing in (\ref{GenExpansion}) to the coincidence limit $x=x'$ we obtain the following diagonal heat kernel expansion
\begin{equation} \label{HKexp}
\hat K_F(\tau|x,x) = \sum\limits_{m=0}^\infty \tau^\frac{m-d/2}{M} \hat E_{2m}(F|x),
\end{equation}
where the Gilkey-Seeley coefficients $\hat E_{2m}(F|x)$ equal
\begin{equation}\label{E2m}
\hat E_{2m}(F|x) = \frac{1}{(4\pi)^{d/2}} \sum\limits_{k=0}^{2m}
\frac{\Gamma\left(k + \frac{d/2-m}{M}\right)}{M\Gamma\left(M k+\frac{d}2-m\right)}\; [\,\hat a_{m,k}(F)\,],
\end{equation}
due to the particular value $\calE_{\nu,\alpha}(0)=\Gamma(\alpha/\nu)/\nu\Gamma(\alpha)$ \cite{Wach3}. Here $[\,\hat a_{m,k}(F)\,]$ follows from the expression (\ref{amk}) where in view of (\ref{NonvanishingCoeff}) the upper summation limit for $l$ is the integer part $\lfloor m-k/2\rfloor$.

%%%%%%%%%%%%%%%%%%%%%%%%%%%%%%%%%%%%%
\section{Perturbation theory} \label{PertTheory}
%%%%%%%%%%%%%%%%%%%%%%%%%%%%%%%%%%%%%

Here we derive the above results. The basic idea of derivation is rather simple. As shown in \cite{Wach3}, the off-diagonal heat kernel expansion for an arbitrary $M$-th power $\hat H^M(\nabla)$ of the minimal second order operator (\ref{minimal2}) can be obtained from its Schwinger-DeWitt expansion (\ref{HeatKernel0}) by the procedure of term-by-term direct/inverse Mellin transform. The result consists in the replacement of the original exponential function by a sequence of generalized ones and by taking relevant fractional powers of the proper time,
\begin{multline} \label{bbKRazlLapl}
\hat K_{H^M}(\tau| x,x') = \frac{\Delta^{1/2}(x,x')}{(4\pi\tau^{1/M})^{d/2}}\sum\limits_{l=0}^\infty \tau^{l/M} \\
\times \calE_{M, d/2-l}\Big(-\frac{\sigma(x,x')}{2\tau^{1/M}}\Big)\,\hat a_l(H|x,x').
\end{multline}
This equation is a particular case of expansions considered in \cite{Wach3}, but for completeness we present its derivation in Appendix \ref{MellinTrick}.

To circumvent the problem that not every minimal operator of order $2M$ can be represented as an $M$-th power of the minimal second order one, we can use the decomposition (\ref{PowerDecomposition}) and consider the operator $\hat H^M(\nabla)$ as an ``unperturbed part'', and the remainder operator $\hat W(\nabla)$ of a positive background dimensionality as a ``perturbation''. To develop such a perturbation theory in powers of the operator $\hat W(\nabla)$ we choose the ansatz
\begin{equation} \label{HeatKernelPert}
\exp\big(\!-\tau \hat F\big) = \hat U_\tau \;\exp\big(\!-\tau \hat H^M\big),
\end{equation}
where $\hat U_\tau$ can be considered as the deformation operator of the heat kernel $K_{H^M}(\tau|x,x')$ into the original heat kernel $\hat K_{F}(\tau|x,x')$. Substituting this ansatz into the heat equation for $\hat K_F(\tau|x,x')$ with the operator $\hat F(\nabla)$ decomposed according to Eq.(\ref{PowerDecomposition}) we obtain the heat equation for $\hat U_\tau$ with the perturbation operator $\hat W_\tau $ in the ``interaction picture'' representation
\begin{align}
&\frac\partial{\partial\tau}\hat U_\tau = -\hat U_\tau \hat W_\tau, \quad
\hat W_\tau = e^{-\tau\hat H^M} \hat W\, e^{\tau\hat H^M}. \label{WIntPic}
\end{align}
The solution is given by the anti-chronologically ordered exponent
\begin{align}
\begin{aligned}
\hat U_\tau &= \bar T\exp\Big(-\int\limits_0^\tau dt\;\hat W_t\Big)\\
&= \sum\limits_{n=0}^\infty (-1)^n\,\int\limits_0^\tau dt_n\int\limits_0^{t_{n}}dt_{n-1}\cdots\!\!
\int\limits_0^{t_2}dt_1\; \hat W_{t_1}\cdots\hat W_{t_n} \label{Texp3}
\end{aligned}
\end{align}
(the anti-chronological nature is, of course, explained by the fact that the interaction Hamiltonian $\hat W_\tau $ in (\ref{WIntPic}) stands to the right of $\hat U_\tau$).

Each of $\hat W_t$ in (\ref{Texp3}) can be expanded in the proper time parameter $t$ by using the well-known relation
\begin{equation} \label{ExpComm}
e^{-tB}\; A\; e^{tB} = \sum\limits_{k=0}^\infty \frac{t^k}{k!}\; \lb A,B\rb_k,
\end{equation}
where the brackets $\lb\ldots,\ldots\rb_k$ denote the multiple nested commutator \eqref{NestedComm}. This leads to the expansion
\begin{equation} \label{WUexpansion}
\hat W_\tau = \sum\limits_{k=0}^\infty \frac{\tau^k}{k!}\; \hat V_k(\nabla),
\end{equation}
with the coefficients $\hat V_k(\nabla)= \lb\hat W(\nabla), \hat H^M\rb_k$ (see Eq.\eqref{Vk}). Growing number of commutators in $\hat V_k(\nabla)$ for growing $k$ implies the growth in background dimensionality because every commutator of $\hat W(\nabla)$ with $\hat H^M(\nabla)$ gives rise either to an extra power of the curvature (generated by the commutator of covariant derivatives) or the covariant derivative of the spacetime dependent coefficients in both operators. Purely matrix commutators in the vector space of hat indices also increase this background dimensionality due to a positive dimension of commuted objects. (The exception from this rule is $\Box^M$-part of $\hat H^M$, which has zero background dimensionality, $dim\,\Box^M=0$, but this part is proportional to the unit matrix, $\Box^M\propto\hat 1$, and the matrix commutation problem does not arise.) This explains why the background dimensionality of $\hat V_k(\nabla)$ is linearly growing with $k$, $dim\,\hat V_k(\nabla)=k+1$, so that the expansion (\ref{WUexpansion}) in powers of the proper time turns out to be sufficient for obtaining the needed gradient and curvature expansion of the heat kernel.

Now, substituting the expansions \eqref{WUexpansion} into (\ref{Texp3}) and taking multiple proper time integrals
\begin{equation}
\int\limits_0^\tau dt_n\int\limits_0^{t_{n}}dt_{n-1}\cdots\!\!
\int\limits_0^{t_2}dt_1\; t_1^{k_1}\cdots t_n^{k_n}=\frac{\tau^{n+|\bm{k}|}}{k_1!\cdots k_n!\,c(\bm{k})}
\end{equation}
we obtain the expansion of the operator $\hat U_\tau$ with the coefficients $\hat U_k(\nabla)$ given by Eqs.\eqref{EvOpExp}-\eqref{c(k)},
\begin{equation} \label{Uexpansion}
\hat U_\tau(\nabla) = \sum\limits_{k=0}^\infty \tau^k \; \hat U_k(\nabla).
\end{equation}

According to Eq.(\ref{HeatKernelPert}) the last step of our derivation consists in the action of the obtained operator $\hat U_\tau(\nabla_x)$ on the heat kernel $\hat K_{H^M}(\tau| x,x')$, which is given by the expansion (\ref{bbKRazlLapl}). As a result the differential operators $\hat U_k(\nabla_x)$ will start acting not only on the Schwinger-DeWitt coefficients $\Delta^{1/2}(x,x')\,\hat a_m(H|x,x')$ but also on the argument $x$ in generalized exponential functions $\calE_{M,\alpha}\big(-\sigma(x,x')/2\tau^{1/M}\big)$ contained in this expansion. Remarkably, these functions can be easily pulled to the left through the operators $\hat U_k(\nabla_x)$, and this procedure results in the sum of their syngifications introduced in Sect.\ref{PertTheory}, $\hat U_k(\nabla_x)\to\{\hat U_k(\nabla_x)\}_n$, along with the set of $n$-dependent shifts of the parameter $\alpha$ of the generalized exponential functions.

This is based on two important identities. The first one is the operator identity applicable to any differentiable function of the Synge world function $f\big(\sigma(x,x')\big)$ and any differential operator $\hat O(\nabla)$ of a finite order $N$,
\begin{equation} \label{CommЕ}
\hat O(\nabla) f(\sigma)
= \sum\limits_{n=0}^N (-2)^n\frac{d^n f(\sigma)}{d\sigma^n}\big\{\hat O(\nabla)\big\}_n.
\end{equation}
This identity should be understood as the operator relation when acting to the right on any test function. Its proof is given below in the Appendix \ref{Identity}. The second identity is the following remarkable differentiation rule for generalized exponential functions
\newpage
\begin{equation} \label{DiffRuleE}
\frac{d^n}{dz^n} \calE_{M,\alpha}(z) = \calE_{M,\alpha+n}(z),
\end{equation}
which follows directly from the definition \eqref{calE}.

Thus, bearing in mind the order $(2M-1)k$ of the differential operator $\hat U_k(\nabla)$ we use these identities with the choice of the function $f(\sigma)=\calE_{M, d/2-l}(-\sigma/2\tau^{1/M})$ and $N=(2M-1)k$ in the composition $\hat U_\tau(\nabla_x)\hat K_{H^M}(\tau| x,x')$ of two operator expansions (\ref{Uexpansion}) and (\ref{bbKRazlLapl}). Collecting together the coefficients of resulting fractional powers of $\tau$ and relevant generalized exponential functions we get
\begin{widetext}
\begin{multline}
\hat K_F(\tau|x,x') = \frac{1}{(4\pi\tau^{1/M})^{d/2}} \sum\limits_{l=0}^\infty \sum\limits_{k=0}^\infty \sum\limits_{n=0}^{(2M-1)k} \tau^{k+\frac{l-n}{M}} \calE_{M,\frac{d}{2}-l+n}\left(-\frac{\sigma}{2\tau^{1/M}}\right) \left\{\hat U_k(\nabla)\right\}_n \Delta^{1/2}(x,x')\,\hat a_l(H|x,x') \\
= \frac{1}{(4\pi\tau^{1/M})^{d/2}} \sum\limits_{m=-\infty}^\infty \sum\limits_{k\ge K_m}^\infty \tau^\frac{m}{M} \calE_{M,\frac{d}{2}+M k-m}\left(-\frac{\sigma}{2\tau^{1/M}}\right) \hat a_{m,k}(F|x,x'), \;\;\; K_m = \max\Big\{0, -\frac{m}{M-1}\Big\}.
\end{multline}
\end{widetext}
Here, when passing to the second line, we replaced the summation variables $(l,k,n)\to(m,k,l)$, $m=Mk-n+l$, and combined all expressions independent of $\tau$ into a new coefficient $\hat a_{m,k}(F|x,x')$. After separating positive and negative fractional powers of $\tau$ this leads to the advocated above expansion \eqref{GenExpansion} and the expression \eqref{amk} for $\hat a_{m,k}(F|x,x')$. This expression together with  \eqref{EvOpExp} reveals the role of the parameter $k$ in $\hat a_{m,k}(F|x,x')$ as the order of perturbation theory in powers of the operator $\hat W(\nabla)$ having a positive background dimensionality $dim\,\hat W(\nabla)\geq 1$.

The structure of the commutator version expansion \eqref{GenExpansion} coincides with the one obtained by the generalized Fourier transform in \cite{Wach3}. It is the same double functional series in $\tau^{m/M} \calE_{M,d/2+Mk-m}(-\sigma/2\tau^{1/M})$. Note, however, that the limits of summation in this expansion are different from those of \cite{Wach3}.

Finally, we show that all coefficients $\hat a_{m,k}(F|x,x')$ vanish in the coincidence limit for $k>2m$, see Eq.\eqref{NonvanishingCoeff}. Indeed, since $[\sigma_a]=0$ and $[\sigma_{ab}]=g_{ab}$, the operator $\{\hat U_k(\nabla)\}_n$ will give a non-zero contribution in the coincidence limit only if each of $n$ functions $\sigma_a$ contained in it is additionally differentiated at least once. Since the operator $\hat U_k(\nabla)$ is of order $(2M-1)k$, only terms with
\begin{equation}
n=Mk+l-m\le (M-\tfrac{1}{2})k.
\end{equation}
survive. Because the summation index $l$ in (\ref{amk}) is always nonnegative, this condition, in its turn, leads to \eqref{NonvanishingCoeff}.

%%%%%%%%%%%%%%%%%%%%%%%%%%%%%%%%%%%%%%%%%%%%%
\section{Examples of applications}
%%%%%%%%%%%%%%%%%%%%%%%%%%%%%%%%%%%%%%%%%%%%%
Here we consider several applications of the commutator technique, which will serve as its verification because the results will be also checked by the alternative method of universal functional traces \cite{Barvinsky1985,JackOsborn1984}. These applications include the calculation of the UV divergent part of the one-loop effective action in a simple model of fourth-order derivative theory, similar calculations for the operator arising in the vector field sector of the (3+1)-dimensional Ho\v rava gravity model and generic fourth-order operator  in curved spacetime.
In what follows we use the following notations for the Riemann curvature of the metric $g_{ab}$ and the fibre-bundle curvature $\hat\calR_{ab}=\hat\calR^A_{B\,ab}$ -- the commutator of covariant derivatives acting in a fibre bundle of spin-tensor fields $\varphi=\varphi^A(x)$ and the corresponding matrices $\hat X=X^A_B(x)$, $\rm tr$ denoting their matrix trace ${\rm tr}\hat X=X^A_A$,
\begin{align}
[\nabla_a,\nabla_b]\,v^c &= R^c{}_{dab}\,v^d, \\
[\nabla_a,\nabla_b]\,\varphi &= \hat\calR_{ab}\,\varphi, \quad [\nabla_a,\nabla_b]\hat X=\big[\hat\calR_{ab},\hat X\big].
\end{align}

\subsection{UV divergences of the one-loop effective
action of a simple fourth-order derivative theory}

Consider a toy model of a scalar field in four dimensional spacetime with the fourth-order inverse propagator
\be\label{opB}
F(\nabla) = \Box^2+P,
\ee
where $P=P(x)$ is some potential term of the dimensionality $dim P=4$.

On the one hand, the UV divergent part of its effective action can be obtained by expanding the logarithm of this operator under the functional trace and using that $\ln(\Box^2)=2\ln(-\Box)$,
\be\label{UFTmain}
\begin{split}
&\varGamma^{\rm div}=\frac12{\rm Tr}\,\ln \left(\Box^2+P\right)\Big|^{\,\rm div}= {\rm Tr}\,\ln(-\Box)\,\Big|^{\,\rm div}\\
&\quad+ \frac12\sum_{k=1}^\infty \frac{(-1)^{k+1}}k \int d^4x\,\left(P\frac1{\Box^2}\right)^k\delta(x,x')\,\Big|^{\,\rm div}_{\,x'=x}.
\end{split}
\ee
Only the first term and $k=1$ term of the sum over $k$ contain here the UV logarithmic divergences. With the use of the proper time representation and the Schwinger-DeWitt expansion for the operator $-\Box$ the first term contributes within dimensional regularization
\be\label{Trlogbox}
\begin{split}
&\!\!\!{\rm Tr}\,\ln(-\Box)\,\Big|^{\,\rm div}=
- \int d^4x\int_0^\infty \frac{d\tau}\tau e^{\tau \Box}\delta(x,x')\,\Big|^{\,\rm div}_{\,x'=x}\\
&\qquad\qquad= - \frac{\ln L^2}{16\pi^2}\int d^4x\,g^{1/2}(x)\, a_2(-\Box|x,x)\;,
\end{split}
\ee
where the second Schwinger-DeWitt coefficient $ a_2(-\Box|x,x)$ of the operator $-\Box$ is given by a classical result of \cite{DeWitt1965} (see also \cite{Barvinsky1985,Scholarpedia}), and the parameter of the dimensional regularization $\ln L^2=1/\varepsilon$, $\varepsilon = 2-\tfrac{d}2\to 0$, serves to extract the logarithmic UV divergence at the lower limit of the proper time integral
\be
\int_0^\infty \frac{d\tau}{\tau^{1-\varepsilon}}\,\Big|^{\,\text{UV div}}=\frac1\varepsilon=\ln L^2. \label{regularization}
\ee

The second term is calculable by the method of universal functional traces---Schwinger-DeWitt expansion applied to the integral representation of the power of the operator $-\Box$
\be\label{1/boxsquared}
\begin{split}
\frac1{\Box^2}\delta(x,x')\,\Big|^{\,\rm div}_{\,x'=x} &=
\int_0^\infty d\tau\,\tau e^{\tau\Box}\delta(x,x')\,\Big|^{\,\rm div}_{x'=x} \\
&=\frac1{16\pi^2}\ln L^2.
\end{split}
\ee
As a result $\varGamma^{\rm div}$ becomes
\be\label{TrBUFT}
\varGamma^{\rm div}= -\frac{\ln L^2}{16\pi^2}\int d^4x\,g^{1/2}\Big([\,a_2(-\Box)\,] -\frac12 P\Big) \;.
\ee

On the other hand, we apply the heat kernel method directly to the fourth-order operator \eqref{opB} and its coincidence limit expansion (\ref{HKexp})
\be
\begin{split}
\varGamma^{\rm div}&
%=- {\rm Tr} \int_0^\infty \frac{d\tau}{\tau} e^{-\tau F(\nabla)}\delta(x,x')\,\Big|^{\,\rm div} \\
%&
= - \int_0^\infty \frac{d\tau}{\tau}  \int d^4x\,g^{1/2} K_F (\tau|x,x)\,\Big|^{\,\rm div}  \\
&= -  \ln L^2\int d^4x\, g^{1/2} E_4(F|x)\;.
\end{split}
\ee
Here this expansion involves half integer powers of the proper time, $\tau^{1/M}=\tau^{1/2}$, thus automatically accounting for its dimension $-4$ (in contrast to the dimension $-2$ in (\ref{Trlogbox})-(\ref{1/boxsquared})) and doubling the expression (\ref{regularization})
\be
\int_0^\infty \frac{d\tau}{\tau^{1-\varepsilon/2}}\,\Big|^{\,\text{UV div}}=\frac2\varepsilon=2\ln L^2.
\ee

To find $E_4(F|x,x)$ we use the commutator version of the heat kernel expansion. To begin with we identify the minimal operator $H(\nabla)$ with $-\Box$ and the perturbation part $W(\nabla)=P$ becomes an ultralocal potential term. Then $E_4(F|x,x)$ is defined by Eq.(\ref{E2m}) with $M=2$
\be
E_4(F|x) = \frac1{32\pi^2}\sum_{k=0}^4 \,\frac{\Gamma(k)}{\Gamma(2k)}\, a_{2,k}(F|x,x)\;,
\ee
where according to (\ref{amk})
\begin{eqnarray}
\hat a_{2,k}(F|x,x')\,&=&\,\sum\limits_{l=2\delta_{0k}}^{2+k}\{\hat U_k(\nabla_x)\}_{2(k-1)+l}\nonumber\\
&&\quad\times
\Delta^{1/2}(x,x')\,\hat a_l(-\Box|x,x'),
\end{eqnarray}
(${\rm max}\{0,2(1-k)\}=2\delta_{0k}=2$ for $k=0$ and 0 for $k>0$). Therefore
\be
\begin{split}
a_{2,0}(F|x) &= a_2(-\Box|x,x),\quad
a_{2,1}(F|x) =  - P(x)\;,\\
a_{2,k}(F|x)&=0, \quad k\geq 2,
\end{split}
\ee
because in view of Eqs.(\ref{U0})-(\ref{EvOpExp}) $U_0 = 1$, $U_1 = - P$, and all $a_{2,k}(F|x)$ with $k\geq2$ are vanishing since all the operators $U_k$ containing derivatives have background dimensionality higher than 4 (and therefore cannot contribute to $E_4$), while the syngification of any ultralocal operator  $\{U_k\}_n$ with $n>0$ is identically vanishing according to the definition (\ref{synge}).

Thus, $16\pi^2E_4(F|x)=[\,a_2(-\Box)\,]-P/2$, and we get the same answer for $\varGamma^{\rm div}$ as the one obtained by the method of universal functional traces \eqref{TrBUFT}, which confirms the validity of our commutator technique.

\subsection{Vector operator of Ho\v rava gravity theory}
Another example demonstrating the power of the commutator technique is the calculations of UV divergences in projectable Ho\v rava gravity theory \cite{Horava}. In contrast to the previous example, however, this calculation cannot be done manually because of extremely high complexity of the computations which have to be performed with the use of symbolic manipulations in both methods -- either in terms of universal functional traces or with the commutator algebra. We will consider only the vector sector of the theory (the contribution of Faddeev-Popov ghosts and shift functions of this model) and demonstrate that both methods lead to the same result, but clearly observe that the number of calculational stages in the functional traces method and their complexity are much higher than in the commutator approach.

The contribution of the vector sector on a static metric background in this model is given by the 4-dimensional functional determinant of the Lorentz violating operator $-\hat 1\partial_t^2+\hat F(\nabla)$ where $\hat F(\nabla)=F^a_{\:\:b}(\nabla)$ is a purely spatial differential operator in spatial covariant derivatives, $a, b=1,2,3$, which in a particular gauge is a minimal 6-th order operator of the form \cite{BarvinskyKurov2022}
\be\label{F}
\begin{aligned}
&F^a_{\:\:b}(\nabla)=(-\Box)^3 \delta^a_b
-\Box^2\nabla_b\nabla^a\\
&+\frac{1-2\lambda}{2(1-\lambda)}\nabla^a\Box\nabla^c\nabla_b\nabla_c
+\frac{1-2\lambda}{2(1-\lambda)}\nabla^a\Box\nabla_b\Box\\
 &+2\lambda \Box^2\nabla^a\nabla_b
 - \frac{\lambda(1-2\lambda)}{1-\lambda}\nabla^a \Box^2\nabla_b\;,
\end{aligned}
\ee
where $\lambda$ is a dimensionless coupling constant. Minimality of this operator is obvious because the sum of coefficients of all terms except the first one is zero, all of them differing only by the ordering of covariant derivatives.

Dimensional reduction allows one to reexpress the spacetime functional determinant in terms of the purely spatial one \cite{BarvinskyKurov2022},
\be\label{Gamma1loop}
\begin{aligned}
\varGamma &= -\frac12 \Tr_4\ln \left( -\hat 1 \p_t^2 + \hat F(\nabla) \right)\;,\\
&=-\frac12\int dt \Tr_3 \sqrt{\hat F(\nabla)}\;.
\end{aligned}
\ee
where the subscript of $\rm Tr$ indicates whether the functional trace is 4-dimensional or 3-dimensional one. Thus the problem reduces to the calculation of the the functional trace of the square root of the 3-dimensional 6-th order vector-field operator.

\subsubsection{The method of universal functional traces}

When converted to explicitly minimal form by commuting all powers of the $\Box$ to the right the operator $F^a_{\:\:b}(\nabla)$ reads
\be\label{split}
\begin{split}
&\hat{F}(\nabla) = (-\Box)^3+\hat W(\nabla), \\
&\hat W(\nabla)= \sum_{a=2}^6 \hat{\cal R}_{(a)}\sum_{a\leq2k\leq 6} \alpha_{a,k} \nabla_1\dots\nabla_{2k-a} (-\Box)^{3-k}\;,
\end{split}
\ee
where the lower derivative part contains as coefficients the tensor quantities $\hat{\cal R}_{(a)}$ built of the curvature tensor and some dimensionless numbers $\alpha_{a,k}$ (we omit their indices and the indices of covariant derivatives). Tensors $\hat{\cal R}_{(a)}$ have the background dimensionality $dim\,\hat{\cal R}_{(a)}=a$.

Extraction of the square root converts the differential operator into the pseudo-differential operator given by an infinite series,
\be
\begin{split}
\sqrt{\hat{F}(\nabla)}&=(-\Box)^{3/2}+ \sum_{a=2}^\infty {\hat{\cal R}}_{(a)} \\
&\times\sum_{k\geq a/2}^{K_a} \tilde\alpha_{a,k} \nabla_1\dots\nabla_{2k-a} \frac{\hat 1}{(-\Box)^{k-3/2}}\;,
\end{split}
\ee
whose new coefficients $\tilde\alpha_{a,k}$ can be obtained by the iterational procedure of the Sylvester type \cite{BarvinskyKurov2022}. Therefore, calculation of the functional trace reduces to the series of universal functional traces of \cite{Barvinsky1985,JackOsborn1984},
\be\label{uft}
\begin{split}
&\!\!\nabla_{a_1}\nabla_{a_2}\dots\nabla_{a_k} \frac{\hat 1}{(-\Box)^\al}\delta(x,x')\,\Big|_{\,x'=x}\\
&\,\,= \frac1{\Gamma(\al)} \int_0^\infty d\tau\, \tau^{\al-1} \nabla_{a_1}\nabla_{a_2}\dots\nabla_{a_k}
\hat K_{-\Box}(\tau|x,x')\,\Big|_{\,x'=x}.
\end{split}
\ee
Their local divergent parts follow from the curvature expansion of the heat kernel and can be tabulated as it was done in \cite{Barvinsky1985,BarvinskyKurov2022}.

Waging through all the above calculational stages and very cumbersome bookkeeping of all contributions (via xAct package of {\em Mathematica} \cite{xAct,Nutma:2013zea}) leads to the final result in terms of five cubic invariants in spatial curvature
\begin{widetext}
\be\label{HGresult}
\begin{aligned}
\varGamma^{\rm div}
= \frac3{32\pi^2} \frac{\ln L^2 }{10080 (1-\lambda)^3}& \int dt\, d^3x\,g^{1/2}
\Big( -(765 - 2869\lambda + 3639\lambda^2 - 1007\lambda^3 - 1752\lambda^4 + 1728\lambda^5 - 512\lambda^6 )  R^3\\
& +(1521 - 8136\lambda + 18409\lambda^2 - 22490\lambda^3 + 15572\lambda^4 - 5632\lambda^5 + 768\lambda^6) R R_{ab} R^{ab}  \\
&-4(873 - 3419 \lambda + 3263\lambda^2 + 2547\lambda^3 - 6672\lambda^4 + 4416 \lambda^5 - 1024\lambda^6) R^a_b R^b_c R^c_a \\
&+(1-\lambda)(313 - 2722\lambda + 6449\lambda^2 - 5760\lambda^3 + 1920\lambda^4)  R\Box R\\
&+ 2(1-\lambda)(617 - 2398\lambda + 4073\lambda^2 - 3056\lambda^3 + 832\lambda^4) R^{ab}\Box R_{ab}\Big).
\end{aligned}
\ee
\end{widetext}
Here the parameter of 3-dimensional regularization is defined in accordance with the dimensionality $-2$ of the parameter $\tau$ in Eq.(\ref{uft}),
\be
\int_0^\infty \frac{d\tau}{\tau^{1-\varepsilon}}\,\Big|^{\text{UV div}}=\ln L^2 = \frac1\varepsilon,\:\:\:\: \varepsilon = \frac{3-d}2\to 0.
\ee

\subsubsection{Commutator algebra method}
In contrast to the complicated iterational procedure of the operator square root extraction, the commutator technique allows us to use directly the proper time integral representation of $\sqrt{F^a_{\:\:b}(\nabla)}$ in terms of its full heat kernel,
\be\label{Gamma1loopHK}
\begin{split}
\!\!\!\varGamma^{\rm div}& = -\frac12  \int dt\, {\rm Tr}_3 \sqrt{\hat F(\nabla)}\,\Big|^{\,\rm div}\\
&= -\frac1{2\Gamma(-1/2)} \int dt \int_0^\infty \frac{d\tau}{\tau^{3/2}} {\rm Tr}_3\,\hat K_F(\tau)\,\Big|^{\,\rm div} \\
&= \frac{\ln L^2}{4\sqrt\pi} \int dt\, d^3x\,g^{1/2}{\rm tr}\,\hat{E}_6(F|x)\,.
\end{split}
\ee
Here $\rm tr$ denotes the matrix trace and we took in account that in dimensional regularization only logarithmic divergences survive. Therefore only one term, $\hat{E}_6(F|x)$, in the heat kernel expansion for the coincidence limit \eqref{HKexp} contributes to the divergent part of the one-loop effective action. Then use the decomposition (\ref{split}) which selects
\be
\hat H=-\Box
\ee
and the perturbation $\hat W(\nabla)$. Then in view of (\ref{E2m}) and (\ref{amk})
\be
\begin{split}
 &\hat E_6(F|x)=  \frac1{8\pi^{3/2}} \sum_{k=0}^6 \frac{\Gamma\left(k-\frac12\right)}{3\Gamma\left(3k-\frac32\right)}\\
&\times\sum_{l=3\delta_{0k}}^{\lfloor3-k/2\rfloor}
 \left[ \big\{\hat{U}_k(\nabla)\big\}_{l+3k-3} \Delta^{1/2}(x,x^\prime)\, \hat{a}_l(H|x,x^\prime) \right] \;.
\end{split}
\ee

Manual manipulations with this expression turns out to be impossible in view of its complexity. For example, the differential operators $\hat V_k$ participating in the construction of $\hat U_k$ and given by multiple commutators (\ref{Vk}) start with 62 terms for $\hat V_0$, 388 terms for $\hat V_1$, 235 terms for $\hat V_2$, etc. So in order to proceed one has to use computer symbolic manipulation facilities. Still there is a big room for analytic simplifications that can be used in the calculation of these structures, which we will briefly explain below.

To begin with note that the functional trace of a single $\hat{V}_k$ is vanishing since every $\hat{V}_k$ for $k\geq1$ is a commutator \eqref{Vk},
\be\label{TrComm}
{\rm Tr}\, \hat{V}_k = {\rm Tr}\, ([ \hat{V}_{k-1},\hat{H}^M])=0, \quad {\rm for} \quad k\geq1.
\ee

That is why in the calculation of operators $\hat{U}_k$, given by the sums of $\hat{V}_k$ monomials \eqref{EvOpExp}, one can omit single $\hat{V}_k$ terms -- these terms will still be present in the local expression for $E_6(F|x)$ but they will represent either matrix commutators or total derivatives that can be disregarded under the spacetime integral of the functional trace (here we are interested only in the bulk part of UV divergences). This is the first type of terms which do not survive after taking of the functional trace. The second type of terms to be disregarded contains syngifications of the form $[\{\nabla_{a_1}\ldots\nabla_{a_n}\}_m]$ with $m>n/2$, i.e. the terms  in which more than one half of their derivatives should be replaced with the world functions $-\sigma^a(x,x^\prime)/2$, so that they vanish in the coincidence limit.

Thus, the relevant remaining $\hat{U}_k$ operators take the form
\be
\begin{aligned}
\hat{U}_0 &= \hat{1}, \quad \hat{U}_1 = - \hat{V}_0, \quad  \hat{U}_2 =
  \frac12\hat{V}_0^2+(\ldots), \\
 \hat{U}_3 &=
 \frac13 \hat{V}_0 \hat{V}_1 + \frac16  \hat{V}_1 \hat{V}_0 - \frac16\hat{V}_0^3+(\ldots) ,\\
 \hat{U}_4 &=
 \frac18 \hat{V}_0 \hat{V}_2 + \frac1{24}  \hat{V}_2 \hat{V}_0 + \frac18\hat{V}_1^2+(\ldots)\: ,\\
 \hat{U}_5 &=0+(\ldots)\;,\quad \hat{U}_6=0+(\ldots)\;,
 \end{aligned}
\ee
where ellipsis imply the terms of the above type which do not contribute to the functional trace.

Thus, with the above simplifications there are five coefficients $\hat{a}_{3,k}(F|x,x)$ which yield nonzero contributions to the bulk part of $ \hat{E}_6(F|x)$,
\be\label{amkF}
\begin{aligned}
&\hat{a}_{3,0}(F|x,x)
=  \hat{a}_3(H|x,x)\;,\\
&\hat{a}_{3,1}(F|x,x)=\sum_{l=0}^2\left[ \big\{\hat{U}_1\big\}_l\, \Delta^{1/2}(x,x^\prime)\, \hat{a}_l(H|x,x^\prime) \right]\;,\\
&\hat{a}_{3,2}(F|x,x) =\sum_{l=0}^1 \Big[ \big\{\hat{U}_2 \big\}_{l+3}\,\Delta^{1/2}(x,x^\prime) \,\hat{a}_l(H|x,x^\prime) \Big]\;,\\
&\hat{a}_{3,3}(F|x,x) = \Big[ \big\{\hat{U}_3\big\}_6 \,\Delta^{1/2}(x,x^\prime)\, \hat{a}_0(H|x,x^\prime) \Big]\;,\\
&\hat{a}_{3,4}(F|x,x) = \left[ \big\{\hat{U}_4\big\}_9 \,\Delta^{1/2}(x,x^\prime)\, \hat{a}_0(H|x,x^\prime) \right]\;,
\end{aligned}
\ee
where the set of operators $\hat U_k$ listed above in terms of $\hat V_k$ should be syngified by the rule (\ref{CurlBrDef})-(\ref{synge}).

The syngification procedure taken in the coincidence limit can be further simplified to make its computer codification more efficient. It reduces to the calculation of coincidence limits
$$
[\{\nabla^x_{a_1}\ldots\nabla^x_{a_n}\}_k\, \Delta^{1/2}(x,x') \hat{a}_l(H|x,x')],
$$
that can be done in two steps -- first consider the coincidence limit of the syngified derivative monomial $[\{\nabla^n\}_k]=[\{\nabla_{a_1}\ldots\nabla_{a_n}\}_k]$ and then apply it to $\Delta^{1/2}(x,x')\, \hat{a}_l(H|x,x')$ followed by the repeated identification $x'=x$. Obviously the result expresses in terms of coincidence limits of numerous covariant derivatives of $\sigma^a(x,x')$, $\Delta^{1/2}(x,x')$ and $\hat{a}_l(H|x,x')$ which are calculable by a known iterational procedure of the Schwinger-DeWitt technique \cite{Barvinsky1985}. In what follows we present several simplifying hints on the calculation of such syngified monomials $[\{\nabla^n\}_k]$. With all the simplifications mentioned above, what one needs for the construction of $\hat{U}_1$ is (indices of derivatives are omitted):
$$
[\{\nabla^4\}_1]\;, \quad [\{\nabla^4\}_2]\;, \quad  [\{\nabla^3\}_1]\;, \quad   [\{\nabla^2 \}_1]\;,
$$
for the construction of $\hat{U}_2$:
$$
[\{\nabla^4 (-\Box^2)\}_3]\;, \:\:\: [\{\nabla^4 (-\Box^2)\}_4]\;,\:\:\:  [\{\nabla^6 \}_3]\;,
$$
while for the construction of $\hat{U}_3$ one needs one set of $[\{\nabla^6 (-\Box)^3\}_6]$ and for $\hat{U}_4$---the set of $[\{\nabla^6 (-\Box)^6\}_9]$.

Some of these operators can be further simplified by using the coincidence limit $[\nabla_a\nabla_b\sigma]=g_{ab}$. The first set consists of $[\{\nabla^{2n}\}_n]$ with the maximum possible number of covariant derivatives replaced by the vector world functions $-\sigma^a(x,x^\prime)/2$. These operators turn out to be ultralocal without derivatives
\be\label{Tracegn}
\left[\{\nabla_{a_1}\ldots\nabla_{a_{2n}}\}_n \right] = \left(-\frac12\right)^n g^{(n)}_{a_1a_2\dots a_{2n}}\,,
\ee
where $g^{(n)}_{a_1a_2\dots a_{2n}}$ is a totally symmetric tensor built of $n$ metric tensor by the iteration scheme $g^{(1)}_{ab} = g_{ab}$, $g^{(n+1)}_{a_1a_2\dots a_{2n+2}} = \sum_{i=2}^{2n+2}\,g_{a_1a_i}\,g^{(n)}_{a_2\ldots \hat a_i\ldots a_{2n+2}}$ ($i$-th hatted index $a_i$ is withdrawn from the tensor and relocated to the metric coefficient $g_{a_1a_i}$). The contraction rule for this tensor
\be
g^{a_1a_2} g^{(n)}_{a_1a_2\dots a_{2n}} = (2n-2+d) g^{(n-1)}_{a_3\dots a_{2n}}\;.
\ee
easily allows one to build the following set of ultralocal operators
\be
\begin{split}
&\left[\{ \nabla_{a_1}\ldots\nabla_{a_{2k}}(-\Box)^{n-k}  \}_n \right]\\
&\qquad= \frac{(-1)^k}{2^n} \frac{(2n-2+d)!!}{(2k-2+d)!!}  g^{(k)}_{a_1a_2\dots a_{2k}}\;.
\end{split}
\ee

Another set of operators $[\{\nabla^{2n+1}\}_n]$ turns out to be of the first order in derivatives
\be
\begin{split}
&\left[\{ \nabla_{a_1}\ldots\nabla_{a_{2n+1}}\}_n\right] =
\left(-\frac12\right)^n \sum_{i=1}^{2n+1} g^{(n)}_{a_1a_2\dots\hat a_i\dots a_{2n+1}}\nabla_{a_i}\;.
\end{split}
\ee
Its contraction with several metrics yields
\be
\begin{split}
&\!\!\!\left[\{ \nabla_{a_1}\ldots\nabla_{a_{2k+1}}(-\Box)^{n-k} \}_n \right] \\
&\quad=  \frac{(-1)^k}{2^n} \frac{(2n+d)!!}{(2k+d)!!}
\sum_{i=1}^{2k+1} g^{(k)}_{a_1a_2\dots\hat a_i\dots a_{2k+1}}\nabla_{a_i}\;.
\end{split}
\ee

More complicated syngified monomials $[\{\nabla^{2n}\}_{n-1}]$ and $[\{\nabla^{2n+1}\}_{n-1}]$ are presented in Appendix \ref{PermTr1}.

Application of the above analytical machinery together with the power of symbolic manipulations of Wolfram {\em Mathematica} \cite{xAct,Nutma:2013zea} shows that the commutator technique result (\ref{Gamma1loopHK}) confirms the expression (\ref{HGresult}) for divergences in Ho\v rava gravity model, which were originally obtained by a lengthy combination of universal functional traces method \cite{Barvinsky1985} and the Sylvester procedure of extracting the operator square root.

\subsection{Generic minimal fourth order operator}
Here we reproduce and correct the results for the heat kernel of a generic fourth order operator
\be\label{F4}
\hat{F}(\nabla) = \Box^2 + \hat{\Omega}^{abc}\nabla_a\nabla_b\nabla_c + \hat{D}^{ab}\nabla_a\nabla_b + \hat{H}^a \nabla_a + \hat{P}\;.
\ee
where the matrix valued coefficients $\hat\varOmega^{abc} = \hat\varOmega^{(abc)}$ and $\hat D^{ab} = \hat D^{(ab)}$ are fully symmetric in spacetime indices. In the case of $\hat\varOmega^{abc}=0$ its low order Gilkey-Seeley coefficients $\hat E_2(x)$ and $\hat E_4(x)$ were calculated under the functional trace sign in
\cite{fradkin1982} and then this result was extended to nonvanishing $\hat\varOmega^{abc}$ in \cite{Barvinsky1985}. Beyond the functional trace but by disregarding the total derivative terms they were later found in \cite{Gusynin1990} again for the case of $\hat\varOmega^{abc}=0$. Their complete calculation  was recently done by the Fourier transform method in \cite{Wach3}, but the result contained several typos in the terms involving a nonzero $\hat\varOmega^{abc}$. Here we correct these results by using the commutator technique.

As in the previous examples, we identify $\hat H(\nabla)$ with $-\Box$, so that the role of perturbation in (\ref{F4}) is played by all four lower derivative terms $\hat W(\nabla)=\hat{\Omega}^{abc}\nabla_a\nabla_b\nabla_c + \hat{D}^{ab}\nabla_a\nabla_b + \hat{H}^a \nabla_a + \hat{P}$. With this decomposition of $\hat F(\nabla)$ we use Eqs.(\ref{amk}), (\ref{U0})-(\ref{EvOpExp}) and (\ref{E2m}) to calculate $\hat E_0(F|x)$, $\hat E_2(F|x)$ and $\hat E_4(F|x)$. Dimensionless $\hat E_0(F|x)$ is obviously a unit matrix with a constant coefficient
\begin{equation}\label{E0}
\hat E_0(F|x) = \frac{\hat 1}{(4\pi)^{d/2}}
\frac{\Gamma\left(\frac{d}4\right)}{2\Gamma\left(\frac{d}2\right)}\; ,
\end{equation}
because $\hat a_{0,0}=\hat 1$.

The calculation of the dimension 2 coefficient $\hat E_2(F|x)$ involves three coincidence limits of the same background dimensionality
\be\label{amkN}
\begin{aligned}
&\big[\,\hat{a}_{1,0}\,\big]
=  \big[\,\hat{a}_1(-\Box)\,\big]\;,\\
&\big[\,\hat{a}_{1,1}\,\big]=\big[\big\{\hat{U}_1\big\}_0 \Delta^{1/2} \hat{a}_0(-\Box)\,\big]\;,\\
&\big[\,\hat{a}_{1,2}\,\big] = \big[\big\{\hat{U}_2 \big\}_{1} \Delta^{1/2} \hat{a}_0(-\Box)\,\big]\;,
\end{aligned}
\ee
where
\be
\begin{aligned}\label{U1-U2}
\hat{U}_1 = - \hat{V}_0=-\hat W, \quad
\hat{U}_2 = -\frac12 \hat{V}_1 +  \frac12 \hat{V}_0^2\;,
 \end{aligned}
\ee
and $\hat W(\nabla)$ in $\hat{V}_0$ and in $\hat{V}_1=[\hat W,\Box^2]$ can be right away truncated to its first two terms because the coefficients $\hat H^a$ and $\hat P$ have dimensions higher than 2 and cannot contribute to $\hat a_{1,k}$ of dimension 2 (note that in view of (\ref{amk}) $dim\,\hat a_{m,k}=2m$).

All this gives after a simple commutator algebra
\begin{equation}
\begin{split}
&\hat E_2(F|x)=\frac{1}{(4\pi)^{d/2}}\frac{\Gamma\big(\frac{d/2-1}{2}\big)}{2\Gamma\lp\frac{d}{2}-1\rp}\lc \frac{1}{2d}\h D+\frac{\h1}{6}R \rc\\
&-\!\frac{1}{(4\pi)^{d/2}}\frac{3\Gamma\big(\frac{d/2-1}{2}\big)}{8d\,\Gamma\lp\frac{d}{2}-1\rp}\!
\lc\!\frac{2\h\O_{abc}\h\O^{abc}+3\h\O_{a}\h\O^{a}}{4(d+4)}+\nb_{a}\h\O^{a}\!\rc,
\end{split}
\label{E2}
\end{equation}
which is the result coinciding with the one obtained in \cite{Gusynin1990} and modified to the case of nonvanishing $\varOmega^{abc}$ in \cite{Wach3}.

The calculation of $\hat E_4(F|x)$ is essentially more complicated. It involves the coincidence limits
\be\label{amkN}
\begin{aligned}
&[\,\hat{a}_{2,0}\,]
=  [\,\hat{a}_2(-\Box)\,]\;,\\
&[\,\hat{a}_{2,1}\,]=\sum_{l=0}^1\big[ \big\{\hat{U}_1\big\}_l \Delta^{1/2} \hat{a}_l(-\Box) \big]\;,\\
&[\,\hat{a}_{2,2}\,] =\sum_{l=0}^1\big[ \big\{\hat{U}_2 \big\}_{l+2} \Delta^{1/2}\hat{a}_l(-\Box) \big]\;,\\
&[\,\hat{a}_{2,3}\,] = \big[ \big\{\hat{U}_3\big\}_4 \Delta^{1/2} \hat{a}_0(-\Box) \big]\;,\\
&[\,\hat{a}_{2,4}\,] =\big[ \big\{\hat{U}_4\big\}_6 \Delta^{1/2} \hat{a}_0(-\Box) \big]\;,
\end{aligned}
\ee
which require in addition to (\ref{U1-U2}) the knowledge of the operators
\begin{widetext}
\be
\begin{aligned}\label{U3-U4}
 \hat{U}_3 &=  - \frac16 \hat{V}_2 + \frac13 \hat{V}_0 \hat{V}_1 + \frac16  \hat{V}_1 \hat{V}_0 - \frac16(\hat{V}_0)^3 ,\\
 \hat{U}_4 &=  - \frac1{24} \hat{V}_3 +  \frac18 \hat{V}_0 \hat{V}_2 + \frac1{24}  \hat{V}_2 \hat{V}_0 + \frac18(\hat{V}_1)^2
 %\\
 %&
-\frac1{24} \hat{V}_1(\hat{V}_0)^2 - \frac1{12} \hat{V}_0\hat{V}_1\hat{V}_0 - \frac18 (\hat{V}_0)^2 \hat{V}_1
 + \frac1{24}(\hat{V}_0)^4\: .
 \end{aligned}
\ee
This time the multiple nested commutators (\ref{Vk}) should include all four coefficients of the perturbation $\hat W(\nabla)$.

Computer symbolic manipulation assembling together the results for (\ref{amkN})-(\ref{U3-U4}) reproduces the answer for the $\hat\varOmega$-independent part of the total $\hat E_4(x)$ obtained in \cite{Wach3},
\begin{align}
&\hat E^{\varOmega=0}_4(x) = \frac{1}{(4\pi)^{d/2}} \frac{\Gamma\left(\frac{d}{4}\right)}{4\Gamma\left(\frac{d}{2}\right)}
\bigg\{(d-2)\Big(\frac{\hat 1}{90}R^2_{abcd}-\frac{\hat 1}{90}R^2_{ab}
%\nonumber\\
%&
+\frac{\hat 1}{36} R^2 + \frac{1}{6} \hat\calR^2_{ab}
+ \frac{\hat 1}{15}\,\Box R\,\Big)
- \frac{1}{3} \hat D^{ab} R_{ab} + \frac{1}{6} \hat D R \nonumber\\
&\qquad\qquad+ \frac{1}{d+2} \Big(\frac{1}{2}\hat D_{ab} \hat D^{ab} + \frac{1}{4}\hat D^2 - \frac{2}{3}(d+1) \nabla_a\nabla_b \hat D^{ab}
%\nonumber\\
%&
+ \frac{1}{6}(d+4) \Box \hat D\Big) - 2\hat P + \nabla_a \hat H^a\bigg\}\;, \label{E4Coeff}
\end{align}
and gives the corrected result for its $\hat\varOmega$-dependent parts of the total $\hat E_4(x)$,
\begin{align}
&\hat E_4(x) =\hat E^{\varOmega=0}_4(x)+\frac{1}{(4\pi)^{d/2}}
\frac{\Gamma\left(\frac{d}{4}\right)}{2\Gamma\left(\frac{d}{2}\right)}
\Big\{ \hat B_\varOmega + \hat B_{\varOmega R} + \hat B_{\varOmega\cal R} + \hat B_{\varOmega DH}\Big\}
\end{align}
These parts are given by the quartic polynomial $ \hat B_\varOmega$ in $\varOmega^{abc}$ and its derivatives,
\begin{equation} \label{next_to_last}
\begin{split}
\hat B_\varOmega&= -\frac{1}{8}(\nabla_a\Box + \Box\nabla_a)\hat\varOmega^a + \frac{d}{8(d+2)} (\nabla_b\nabla_a\nabla^b\hat\varOmega^a + 2\nabla_a\nabla_b\nabla_c\hat\varOmega^{abc})\\
&+ \frac{1}{32(d+2)(d+6)}
\Big\{  (d+8)\; \Big[6(\nabla_a\hat\varOmega^a)^2 + 3(\nabla_a\hat\varOmega_b)^2+  12(\nabla_a\hat\varOmega^{abc})^2 + 2(\nabla_a\hat\varOmega^{bcd})^2 \Big]
- 6(d+4)(\nabla_b\hat\varOmega_c)\nabla_a\hat\varOmega^{abc}\\
&\qquad\qquad\qquad\qquad\quad- 6(3d+16)(\nabla_a\hat\varOmega^{abc})\nabla_b\hat\varOmega_c\Big\}\\
&+ \frac{1}{32(d+2)(d+6)}\nabla_a
\Big\{6(d+4)\Big[\nabla_b(\hat\varOmega^{abc}\hat\varOmega_c)\!+\!(\nabla_b\hat\varOmega_c)\hat\varOmega^{abc}\!+\!2(\nabla^d\hat\varOmega_{bcd})\hat\varOmega^{bca}\!
+\!(\nabla_b\hat\varOmega^b) \hat\varOmega^a\Big]\!+ 6\,(3d+16)\,\hat\varOmega_b\nabla_c\hat\varOmega^{abc}\\
&\qquad\qquad\qquad\qquad\quad-(d+8)\Big[\nabla^a( 2\hat\varOmega_{bcd}\hat\varOmega^{bcd}\! +\!3\hat\varOmega_a\hat\varOmega^a )\! +\!3\hat\varOmega^a\nabla_b\hat\varOmega^b\!
 +\!3\hat\varOmega^b\nabla_b\hat\varOmega^a\!
 + 6(\hat\varOmega^{abc}\nabla^d\hat\varOmega_{dbc}\!
+\!\hat\varOmega_{bcd}\nabla^b\hat\varOmega^{acd})\Big] \Big\} \\
&- \frac{g^{(4)}_{a_1a_2\dots a_8}}{32(d+2)(d+6)} \Big\{\hat\varOmega^{ba_1a_2}\nabla_b(\hat\varOmega^{a_3a_4a_5}\hat\varOmega^{a_6a_7a_8}) + \hat\varOmega^{a_1a_2a_3}\hat\varOmega^{ba_4a_5}\nabla_b\hat\varOmega^{a_6a_7a_8}\Big\}\\
&+ \frac{g^{(5)}_{ba_1a_2\dots a_9}}{384(d+2)(d+6)}[\hat\varOmega^{a_1a_2a_3}\hat\varOmega^{a_4a_5a_6},\nabla^b\hat\varOmega^{a_7a_8a_9}]
+\frac{1}{192(d+2)(d+6)} \nabla^b\Big(g^{(5)}_{ba_1a_2\dots a_9}\hat\varOmega^{a_1a_2a_3}\hat\varOmega^{a_4a_5a_6}\hat\varOmega^{a_7a_8a_9}\!\Big)\\
&+\frac{g^{(6)}_{a_1a_2\dots a_{12}}}{1536(d+2)(d+6)(d+10)}
\hat\varOmega^{a_1a_2a_3}\hat\varOmega^{a_4a_5a_6}\hat\varOmega^{a_7a_8a_9}\hat\varOmega^{a_{10}a_{11}a_{12}},
\end{split}
\end{equation}
and the cross terms of $\hat\varOmega^{abc}$ with Ricci curvature, fibre bundle curvature $\hat{\cal R}_{ab}$ and the coefficients of the operator $\hat D^{ab}$ and $\hat H^a$,
\begin{equation}\label{last}
\begin{split}
&\hat B_{\varOmega R} = -\frac{1}{8} R\nabla_a\hat\varOmega^a + \frac{1}{4}R_{ab}\nabla_c\hat\varOmega^{abc}
+ \frac{1}{32(d+2)} \bigg\{{6R_{ab}\hat\varOmega^{abc}\hat\varOmega_c} + 6R_{ab}\hat\varOmega_c\hat\varOmega^{abc} - 3R\hat\varOmega_a^2 - 2R\hat\varOmega_{abc}^2\\
&\qquad\qquad\qquad\qquad\;\;\;+ 4\hat\varOmega^a\nabla_aR + 8R_{ab}\nabla^a\hat\varOmega^{b}-3\frac{d+4}{d+6}\,
\big(\,
4R_{abcd}\hat\varOmega_e^{\;\;ac}\hat\varOmega^{ebd}-2R_{ab}\hat\varOmega^a_{\;\;cd}\hat\varOmega^{bcd}
-R_{ab}\hat\varOmega^a\hat\varOmega^b \big)\bigg\},\\
%\end{split}
%\end{equation}
%\begin{equation}
&\hat B_{\varOmega\calR} = -\frac{1}{16(d+2)} \bigg\{4(d+4)(\nabla_b\hat\varOmega_a)\hat\calR^{ab} - 8\hat\varOmega_b\nabla_a\hat\calR^{ab} - 6\hat\varOmega_{acd}\hat\varOmega_b{}^{cd}\hat\calR^{ab} - 3\hat\varOmega_a\hat\varOmega_b\hat\calR^{ab}\\
&\qquad\qquad\qquad\qquad\;\;\;-\frac32\frac{d+4}{d+6}\,
\Big(2\,[\,\hat{\cal R}_{ab},\hat\varOmega_{cd}^{\;\;\;\,a}\,]\,\hat\varOmega^{cdb}
+ [\,\hat{\cal R}_{ab},\hat\varOmega^a]\,\hat\varOmega^b\Big) \bigg\},\\
%\end{split}
%\end{equation}
%begin{equation}\label{lastlast}
%\begin{split}
&\hat B_{\varOmega DH} = \frac{3}{8(d+2)} \big(\hat H_a\hat\varOmega^a + \hat\varOmega^a\hat H_a\big)
+ \frac{1}{8(d+2)}\Big\{- 6\hat D_{ab}\nabla_c\hat\varOmega^{abc}-6\hat\varOmega_a\nabla_b\hat D^{ab}- 3 \hat D\nabla_a\hat\varOmega^a
\\
&\qquad\qquad- 2 [\,\nabla_a\hat\varOmega^{abc},\hat D_{bc}\,]
+2[\,\hat\varOmega^{abc},\nabla_a\hat D_{bc}]- 2[\,\nabla_a\hat D^{ab},\hat\varOmega_b\,] + 2[\,\hat D^{ab},\nabla_a\hat\varOmega_b\,] - [\,\nabla_a\hat\varOmega^a,\hat D\,]+[\,\hat\varOmega^a,\nabla_a\hat D\,]\Big\}\\
&\qquad\qquad
- \frac{g^{(4)}_{a_1a_2\dots a_8}}{96(d+2)(d+6)} \big(\hat D^{a_1a_2}\hat\varOmega^{a_3a_4a_5}\hat\varOmega^{a_6a_7a_8} + \hat\varOmega^{a_1a_2a_3}\hat D^{a_4a_5}\hat\varOmega^{a_6a_7a_8} + \hat\varOmega^{a_1a_2a_3}\hat\varOmega^{a_4a_5a_6}\hat D^{a_7a_8}\big).
\end{split}
\end{equation}
\end{widetext}
In addition to corrections of the results of \cite{Wach3} the above expressions (\ref{next_to_last})-(\ref{last}) feature clearly disentangled total derivative and total commutator terms which disappear under the sign of either the functional or matrix traces.

%%%%%%%%%%%%%%%%%%%%%%%%%%%%%%%%%%%%%%%%%%
\section{Discussion and conclusions}
%%%%%%%%%%%%%%%%%%%%%%%%%%%%%%%%%%%%%%%%%%
The advantage of the commutator technique lies in its universality, because it allows  one to avoid auxiliary procedures like a perturbative extraction of the operator square roots or covariant Fourier transforms of \cite{Wach3} which lead to complicated iteration schemes for auxiliary structures. In particular, this technique makes verifiable the calculations for a generic fourth-order operator and allows one to get rid of errors which might be inevitable within less efficient schemes.

Despite transparency of the suggested formalism its implementation even in simple examples of the above type looks very involved and demands the use of symbolic manipulation programs like xAct package of Wolfram {\em Mathematica} \cite{xAct,Nutma:2013zea}. However, the analytic framework of the above commutator algebra does not yet seem to be completely exhausted, because a number of hidden properties of generalized exponential functions, relevant coefficients $\hat a_{m,k}(F|x,x')$ and Synge world function still remain to be revealed and properly understood. In particular, it should be emphasized that it is not yet clear what type of hidden identities between these objects can be responsible for an explicit proof of the equivalence of the commutator technique and original Fourier method of \cite{Wach3}. One of the examples of the puzzling efficiency of commutator version as compared to the Fourier transform is the problem of the range of summation over the index $k$ of $\hat a_{m,k}(F|x,x')$. It looks as follows.

General formalism of Fourier method suggested that it should be bounded from above by $4m$ \cite{Wach3}, though concrete calculations made within this method showed that $\hat a_{m,k}(F|x,x')$ start vanishing above $k=2m$, which is exactly a firm prediction (\ref{NonvanishingCoeff}) of the commutator algebra technique. This indicates to a stronger universality of the commutator technique, but leaves unclear what kind of relations between the objects of the above type might be responsible for that. Possible source of these relations might be the actual independence of the final result from the choice of the minimal operator $\hat H(\nabla)$ in the decomposition (\ref{PowerDecomposition}). Indeed, the choice of the potential term $\hat P(x)$ in $\hat H(\nabla)$ is an arbitrary free element of the formalism, and the freedom in its choice can be used for obtaining certain nontrivial identities involving the coefficients $\hat a_{m,k}(F|x,x')$, generalized exponentials, Synge world function and its derivatives.

Commutator algebra technique for heat kernel seems to be an essential step forward beyond other calculational methods. However, the boundaries of this research direction are not yet pushed to their logical extreme, because there still remains a realm of nonminimal differential operators which might play important role in the search for renormalizable, UV complete theory of gravity \cite{Horava} and other applications in high energy physics, quantum gravity and other areas of theoretical physics. Thus, the heat kernel technique should be extended to the operators with the minimality condition replaced by causality \cite{Barvinsky1985,Scholarpedia} and beyond. Known functorial properties of the heat kernel formalism, already used in the above studies, might give a clue to the solution of this problem, and this is a subject of future research.

\section*{Acknowledgements}
The authors are indebted to A.Kalugin for critical remarks and helpful verification of heat kernel calculations. A.O.B. is especially grateful for fruitful discussions and correspondence with John Donoghue, Hugh Osborn and Roberto Percacci. This work was supported by the Russian Science Foundation grant No 23-12-00051, https://rscf.ru/en/project/23-12-00051/.

\appendix

\section{The heat kernel expansion for a power of the minimal second order operator} \label{MellinTrick}
%%%%%%%%%%%%%%%%%%%%%%%%%%%%%%%%%%%%%%%%%%%%%%

The derivation of the expansion \eqref{bbKRazlLapl} is based on two facts. The first one is that the heat kernel $\hat K_H(\tau|x,x')$ and the Green's function $\hat G_{H^s}(x,x') = \hat H^{-s}(\nabla)\delta(x,x')g^{-1/2}(x')$ of the complex power $s$ of the operator $H(\nabla)$ are related by the direct and inverse Mellin transforms, \cite{Wach2,Wach3}
\begin{align}
\hat G_{H^s}(x,x') &= \frac{1}{\Gamma(s)} \int\limits_0^\infty d\tau\, \tau^{s-1} \hat K_H(\tau|x,x'), \label{FnGr} \\
\hat K_H(\tau|x,x') &= \frac{1}{2\pi i} \int\limits_{w-i\infty}^{w+i\infty} ds\, \tau^{-s}\, \Gamma(s)\,\hat G_{H^s}(x,x'), \label{KFFs}
\end{align}
where $w$ is some positive real number.

The second fact is that the generalized exponential functions $\calE_{\nu,\alpha}(z)$ (\ref{calE}) have a very simple representation in terms of the Mellin-Barnes integral also related to their inverse and direct Mellin transforms,
\begin{align}
&\calE_{\nu,\alpha}(-z) = \frac{1}{2\pi i} \int\limits_C ds\,
\frac{\Gamma(s)\Gamma\left(\frac{\alpha-s}{\nu}\right)}{\nu\Gamma(\alpha-s)}\,z^{-s}, \label{InvMellin} \\
&\int\limits_0^\infty dz\, z^{s-1} \calE_{\nu,\alpha}(-z)=
\frac{\Gamma(s)\Gamma\left(\frac{\alpha-s}{\nu}\right)}{\nu\Gamma(\alpha-s)}. \label{MellinCalE}
\end{align}
Here $C$ is such a deformation of the integration contour in Eq.(\ref{KFFs}) that it separates the poles of the gamma function $\Gamma(s)$ running to the left of the complex plane of $s$ and the poles of the function $\Gamma((\alpha-s)/\nu)$ running to the right.

Then, substituting the Scwinger-DeWitt expansion (\ref{HeatKernel0}) into the direct Mellin transformation \eqref{FnGr} and integrating the series term by term, we obtain
\begin{equation} \label{Fs}
\hat G_{H^s}(x,x') = \Delta^{1/2}(x,x')\sum\limits_{m=0}^\infty \bbG_m(s, \sigma)\,\hat a_m(H|x,x'),
\end{equation}
where the set of basis Green's functions $\bbG_m(s, \sigma)$ is given by easily calculable integrals
\begin{align}
\bbG_m(s,\sigma) &= \frac{1}{\Gamma(s)} \int\limits_0^\infty d\tau\,
\frac{\tau^{s-1+m}}{(4\pi \tau)^{d/2}}\exp\left(-\frac{\sigma}{2\tau}\right) \nonumber \\
&= \frac{\Gamma(d/2-m-s)}{(4\pi)^{d/2}\Gamma(s)} \left(\frac{\sigma}{2}\right)^{s+m-\frac{d}{2}}.
\end{align}

Now, in turn, we substitute the expansion \eqref{Fs} into the inverse Mellin transformation \eqref{KFFs} for the operator power $\hat H^p$. Integrating the series term by term once again, we find the following heat kernel expansion
\begin{equation} \label{PowerKernelExpansion}
\hat K_{H^p}(\tau|x,x') = \Delta^{1/2}(x,x')\sum\limits_{m=0}^\infty \bbK_m^{(p,d)}(\tau,\sigma)\,\hat a_m(H|x,x'),
\end{equation}
where the new set of basis kernels $\bbK_m^{(p,d)}(\tau,\sigma)$ is given by the integrals
\begin{align}
\bbK_m^{(p,d)}(\tau,\sigma) &= \frac{1}{2\pi i} \int\limits_{w-i\infty}^{w+i\infty} ds\, \tau^{-s} \Gamma(s)\,\bbG_m(ps, \sigma) \nonumber \\
&= \frac{\tau^{m/p}}{(4\pi\tau^{1/p})^{d/2}}\, \calE_{p,\frac{d}{2}-m}\left(-\frac{\sigma}{2\tau^{1/p}}\right), \label{PowerBasisKernels}
\end{align}
which are again easily calculable by changing the integration variable $s\to t=d/2-m-ps$ and using the  Mellin-Barnes representation \eqref{InvMellin} of the generalized exponential function with $\nu=p$, $\alpha=d/2-m$ and $z = \sigma/2\tau^{1/p}$.

So raising a minimal second order operator $\hat H(\nabla)$ to a power $p$ simply amounts to replacing the basis kernels of the Schwinger-DeWitt expansion $\tau^m e^{-\sigma/2\tau}/(4\pi\tau)^{d/2}$ by $\bbK_m^{(p,d)}(\tau,\sigma)$ \eqref{PowerBasisKernels}, while keeping its off-diagonal coefficients $\hat a_m(H|x,x')$ unchanged. We call this remarkable property ``the generalized functoriality''. For integer $p=M$ the formulae \eqref{PowerKernelExpansion}-\eqref{PowerBasisKernels} yield the expansion \eqref{bbKRazlLapl}.

\section{Syngification identity}\label{Identity}
To prove (\ref{CommЕ}) it is sufficient to consider it for the case of the operator represented by a single monomial in covariant derivatives of some order $N$, $\nabla_{a_1}\cdots\nabla_{a_N}$. Then, by linearity it will hold for a generic covariant differential operator.

The differentiable function $f(\sigma)$ in (\ref{CommЕ}) can be represented by a Taylor series in the form
\begin{equation}
f(\sigma)=\sum\limits_{n=0}^\infty\frac1{n!}f^{(n)}(0)\,\sigma^n=
\sum\limits_{n=0}^\infty\frac1{n!}f^{(n)}(0)\,\frac{d^n}{dk^n}\,e^{\sigma k}\,\bigg|_{\,k=0},
\end{equation}
so that the operator $\nabla_{a_1}^x\cdots\nabla_{a_N}^x f\big(\sigma(x,x')\big)$ when it is acting to the right on some function of $x$
can be rewritten as
\begin{widetext}
\begin{eqnarray}
\nabla_{a_1}^x\cdots\nabla_{a_N}^x f\big(\sigma(x,x')\big)\;&&=
\sum\limits_{n=0}^\infty\frac1{n!}f^{(n)}(0)\,\Big(e^{-k\sigma}\frac{d^n}{dk^n}\,
\nabla_{a_1}\cdots\nabla_{a_N} e^{k\sigma}\Big)\,\bigg|_{\,k=0}\nonumber\\
&&=\sum\limits_{n=0}^\infty\frac1{n!}f^{(n)}(0)\,\left(\frac{d}{dk}+\sigma\right)^n\,
(\nabla_{a_1}+k\sigma_{a_1})\cdots(\nabla_{a_N}+k\sigma_{a_N})\,\bigg|_{\,k=0}\nonumber\\
&&=\sum\limits_{n=0}^\infty\sum\limits_{l=0}^n \frac{f^{(n)}(0)}{l!(n-l)!}\,\sigma^l\,\frac{d^{n-l}}{dk^{n-l}}
(\nabla_{a_1}+k\sigma_{a_1})\cdots(\nabla_{a_N}+k\sigma_{a_N})\,\bigg|_{\,k=0}\nonumber\\
&&=\sum\limits_{p=0}^N\bigg(\sum\limits_{l=0}^\infty \frac{1}{l!}f^{(l+p)}(0)\,\sigma^l\bigg)\,
\frac1{p!}\frac{d^p}{dk^p}
(\nabla_{a_1}+k\sigma_{a_1})\cdots(\nabla_{a_N}+k\sigma_{a_N})\,\bigg|_{\,k=0}
\end{eqnarray}
It is not hard to recognize in big parenthesis the $p$-th order derivative of $f(\sigma)$, while the rest according to the definition (\ref{synge}) of syngification is just $(-2)^p\{\nabla_{a_1}\cdots\nabla_{a_N}\}_p$. Therefore we get the equation which finally proves Eq.(\ref{CommЕ}),
\begin{equation}
\nabla_{a_1}\cdots\nabla_{a_N} f(\sigma)=\sum\limits_{p=0}^N (-2)^p\,\frac{d^p f(\sigma)}{d\sigma^p}\,\{\nabla_{a_1}\cdots\nabla_{a_N}\}_p.
\end{equation}

\section{Syngifications of the covariant derivative monomials}\label{PermTr1}

We provide the most general expression for the trace of the form $[\{\nabla^{2n}\}_{n-1}]$
\be\label{2nn-1}
\begin{aligned}
\big[\{\nabla_{a_1}\nabla_{a_2}\dots \nabla_{a_{2n}} \}_{n-1}\big]
=&\left(-\frac12\right)^{n-1} \bigg(\sum_{k<l} g^{(n-1)}_{a_1a_2\dots\hat a_k\hat a_l\dots a_{2n}}\,\nabla_{a_k}\nabla_{a_l}\\
&+\sum_{k<l<p<q} g^{(n-2)}_{a_1a_2\dots\hat a_k\hat a_l\hat a_p\hat a_q\dots a_{2n}}\,\big[\nabla_{a_k}\nabla_{a_l}\nabla_{a_p}\nabla_{a_q}\sigma\big]  \bigg)\;,\quad k,l,p,q = 1,2,\dots,2n\;.
\end{aligned}
\ee
The first sum is taken over all terms with $k$-th and $l$-th indices, $k<l$, which are withdrawn from the completely symmetric tensor $g^{(n)}_{a_1a_2\dots a_{2n}}$ and relocated to the two covariant derivatives $\nabla_{a_k}\nabla_{a_l}$, the hat over index indicating that this index is absent in the symmetric tensor $g^{(n-1)}_{a_1a_2\dots\hat a_k\hat a_l\dots a_{2n}}$ whose rank now is $2(n-1)$. In the second sum a similar procedure concerns four indices relocated to the fourth order covariant derivative of the world function with the following transition to its coincidence limit. Obviously the ordering of covariant derivatives in both sums is preserved the same as in the left hand side of the equation. We demonstrate the implementation of the  described above rules  on a simple example
\begin{eqnarray}
\big[\{\nabla_a\nabla_b\nabla_c\nabla_d\}_1 \big]=\,&&-\frac12\big(g_{cd}\nabla_a\nabla_b +g_{bd}\nabla_a\nabla_c+g_{bc}\nabla_a\nabla_d+g_{ac}\nabla_b\nabla_d+g_{ad}\nabla_b\nabla_c+g_{ab}\nabla_c\nabla_d \big)\nonumber \\
&&-\frac12\big[\nabla_a\nabla_b\nabla_c\nabla_d\sigma(x,x^\prime)\big].
\end{eqnarray}

Other examples of traces with metric contractions are
%\begin{widetext}
\be
\begin{aligned}
&\big[\{\nabla_a\nabla_b  \}_0\big] = \nabla_a\nabla_b,\\
&\big[\{\nabla_a\nabla_b \Box \}_1\big] = -\frac12\Big( g_{ab}\Box + (d+2)\nabla_a\nabla_b + 2\nabla_b\nabla_a
- \tfrac23 R_{ij}\Big),\\
&\big[\{\nabla_a\nabla_b \Box^n \}_n\big] =\Big(\!-\frac12\Big)^n (d+4)(d+6)\dots(d+2n-2)\Big( n(d+2n)g_{ab}\Box + \big(d^2 + 2 (n+1)(d+n)\big)\nabla_a\nabla_b\\
&\qquad\qquad\qquad+ 2n(d+n+1) \nabla_b\nabla_a \Big)+ O[{\cal R}],\quad n\geq 2.
\end{aligned}
\ee
%\end{widetext}
For $n=2$ the last equation holds with the overall coefficient $(d+4)(d+6)\dots(d+2n-2)$ replaced by 1.

For generality we also consider syngifications of the form $[\{\nabla^{2n+1}\}_{n-1}]$.  The most general one in this case has the form
\be\label{2n+1n-1}
\begin{aligned}
&\big[\{\nabla_{a_1}\nabla_{a_2}\dots \nabla_{a_{2n+1}} \}_{n-1}\big]=\left(-\frac12\right)^{n-1} \bigg(\sum_{k<l<p} g^{(n-1)}_{a_1a_2\dots\hat a_k\hat a_l\hat a_p\dots a_{2n+1}}\,\nabla_{a_k}\nabla_{a_l}\nabla_{a_p}\\
&+\sum_{k<l<p<q} g^{(n-2)}_{a_1a_2\dots\hat a_k\hat a_l\hat a_p\hat a_q\hat a_r\dots a_{2n+1}}
\big[\nabla_{a_k}\nabla_{a_l}\nabla_{a_p}\nabla_{a_q}\sigma(x,x^\prime)\big] \,\nabla_{a_r}\\
&+\sum_{k<l<p<q<r}  g^{(n-2)}_{a_1a_2\dots\hat a_k\hat a_l\hat a_p\hat a_q\hat a_r\dots a_{2n+1}}\,\left[\nabla_{a_k}\nabla_{a_l}\nabla_{a_p}\nabla_{a_q}\nabla_{a_r}
\sigma(x,x^\prime)\right]\bigg)\;,\quad k,l,p,q,r = 1,2,\dots,2n+1\;.
\end{aligned}
\ee
where the sums are taken over all relocations of three, four and five indices of derivatives which retain their original order.
\end{widetext}
\bibliography{Wachowski2408}

\begin{thebibliography}{36}%
\makeatletter
\providecommand \@ifxundefined [1]{%
 \@ifx{#1\undefined}
}%
\providecommand \@ifnum [1]{%
 \ifnum #1\expandafter \@firstoftwo
 \else \expandafter \@secondoftwo
 \fi
}%
\providecommand \@ifx [1]{%
 \ifx #1\expandafter \@firstoftwo
 \else \expandafter \@secondoftwo
 \fi
}%
\providecommand \natexlab [1]{#1}%
\providecommand \enquote  [1]{``#1''}%
\providecommand \bibnamefont  [1]{#1}%
\providecommand \bibfnamefont [1]{#1}%
\providecommand \citenamefont [1]{#1}%
\providecommand \href@noop [0]{\@secondoftwo}%
\providecommand \href [0]{\begingroup \@sanitize@url \@href}%
\providecommand \@href[1]{\@@startlink{#1}\@@href}%
\providecommand \@@href[1]{\endgroup#1\@@endlink}%
\providecommand \@sanitize@url [0]{\catcode `\\12\catcode `\$12\catcode
  `\&12\catcode `\#12\catcode `\^12\catcode `\_12\catcode `\%12\relax}%
\providecommand \@@startlink[1]{}%
\providecommand \@@endlink[0]{}%
\providecommand \url  [0]{\begingroup\@sanitize@url \@url }%
\providecommand \@url [1]{\endgroup\@href {#1}{\urlprefix }}%
\providecommand \urlprefix  [0]{URL }%
\providecommand \Eprint [0]{\href }%
\providecommand \doibase [0]{https://doi.org/}%
\providecommand \selectlanguage [0]{\@gobble}%
\providecommand \bibinfo  [0]{\@secondoftwo}%
\providecommand \bibfield  [0]{\@secondoftwo}%
\providecommand \translation [1]{[#1]}%
\providecommand \BibitemOpen [0]{}%
\providecommand \bibitemStop [0]{}%
\providecommand \bibitemNoStop [0]{.\EOS\space}%
\providecommand \EOS [0]{\spacefactor3000\relax}%
\providecommand \BibitemShut  [1]{\csname bibitem#1\endcsname}%
\let\auto@bib@innerbib\@empty
%</preamble>
\bibitem [{\citenamefont {Schwinger}(1951)}]{Schwinger}%
  \BibitemOpen
  \bibfield  {author} {\bibinfo {author} {\bibfnamefont {J.}~\bibnamefont
  {Schwinger}},\ }\bibfield  {title} {\bibinfo {title} {{On gauge invariance
  and vacuum polarization}},\ }\href {https://doi.org/10.1103/PhysRev.82.664}
  {\bibfield  {journal} {\bibinfo  {journal} {Phys. Rev.}\ }\textbf {\bibinfo
  {volume} {82}},\ \bibinfo {pages} {664} (\bibinfo {year} {1951})}\BibitemShut
  {NoStop}%
\bibitem [{\citenamefont {DeWitt}(1965)}]{DeWitt1965}%
  \BibitemOpen
  \bibfield  {author} {\bibinfo {author} {\bibfnamefont {B.~S.}\ \bibnamefont
  {DeWitt}},\ }\href@noop {} {\emph {\bibinfo {title} {{Dynamical Theory of
  Groups and Fields}}}}\ (\bibinfo  {publisher} {Gordon and Breach},\ \bibinfo
  {address} {New York},\ \bibinfo {year} {1965})\BibitemShut {NoStop}%
\bibitem [{\citenamefont {Barvinsky}\ and\ \citenamefont
  {Vilkovisky}(1985)}]{Barvinsky1985}%
  \BibitemOpen
  \bibfield  {author} {\bibinfo {author} {\bibfnamefont {A.~O.}\ \bibnamefont
  {Barvinsky}}\ and\ \bibinfo {author} {\bibfnamefont {G.~A.}\ \bibnamefont
  {Vilkovisky}},\ }\bibfield  {title} {\bibinfo {title} {{The generalized
  Schwinger--DeWitt technique in gauge theories and quantum gravity}},\ }\href
  {https://doi.org/10.1016/0370-1573(85)90148-6} {\bibfield  {journal}
  {\bibinfo  {journal} {Phys. Rep.}\ }\textbf {\bibinfo {volume} {119}},\
  \bibinfo {pages} {1} (\bibinfo {year} {1985})}\BibitemShut {NoStop}%
\bibitem [{\citenamefont {Barvinsky}(2015)}]{Scholarpedia}%
  \BibitemOpen
  \bibfield  {author} {\bibinfo {author} {\bibfnamefont {A.~O.}\ \bibnamefont
  {Barvinsky}},\ }\bibfield  {title} {\bibinfo {title} {{Heat kernel expansion
  in the background field formalism}}} (\bibinfo {year} {2015}),\ \bibinfo
  {note}
  {\href{http://www.scholarpedia.org/article/Heat_kernel_expansion_in_the_background_field_formalism}{{S}cholarpedia,
  10(6):31644}}\BibitemShut {NoStop}%
\bibitem [{\citenamefont {Stelle}(1977)}]{Stelle}%
  \BibitemOpen
  \bibfield  {author} {\bibinfo {author} {\bibfnamefont {K.~S.}\ \bibnamefont
  {Stelle}},\ }\bibfield  {title} {\bibinfo {title} {{Renormalization of
  higher-derivative quantum gravity}},\ }\href
  {https://doi.org/10.1103/PhysRevD.16.953} {\bibfield  {journal} {\bibinfo
  {journal} {Phys. Rev.}\ }\textbf {\bibinfo {volume} {D16}},\ \bibinfo {pages}
  {953} (\bibinfo {year} {1977})}\BibitemShut {NoStop}%
\bibitem [{\citenamefont {Fradkin}\ and\ \citenamefont
  {Tseytlin}(1982)}]{fradkin1982}%
  \BibitemOpen
  \bibfield  {author} {\bibinfo {author} {\bibfnamefont {E.~S.}\ \bibnamefont
  {Fradkin}}\ and\ \bibinfo {author} {\bibfnamefont {A.~A.}\ \bibnamefont
  {Tseytlin}},\ }\bibfield  {title} {\bibinfo {title} {{Renormalizable
  asymptotically free quantum theory of gravity}},\ }\href
  {https://doi.org/10.1016/0550-3213(82)90444-8} {\bibfield  {journal}
  {\bibinfo  {journal} {Nucl. Phys. B}\ }\textbf {\bibinfo {volume} {B201}},\
  \bibinfo {pages} {469} (\bibinfo {year} {1982})}\BibitemShut {NoStop}%
\bibitem [{\citenamefont {Avramidy}\ and\ \citenamefont
  {Barvinsky}(1985)}]{Avramidy1985}%
  \BibitemOpen
  \bibfield  {author} {\bibinfo {author} {\bibfnamefont {I.~G.}\ \bibnamefont
  {Avramidy}}\ and\ \bibinfo {author} {\bibfnamefont {A.~O.}\ \bibnamefont
  {Barvinsky}},\ }\bibfield  {title} {\bibinfo {title} {{Asymptotic freedom in
  higher-derivative quantum gravity}},\ }\href
  {https://doi.org/10.1016/0370-2693(85)90248-5} {\bibfield  {journal}
  {\bibinfo  {journal} {Phys. Lett. B}\ }\textbf {\bibinfo {volume} {159}},\
  \bibinfo {pages} {269} (\bibinfo {year} {1985})}\BibitemShut {NoStop}%
\bibitem [{\citenamefont {Codello}\ and\ \citenamefont
  {Percacci}(2006)}]{Codello:2006in}%
  \BibitemOpen
  \bibfield  {author} {\bibinfo {author} {\bibfnamefont {A.}~\bibnamefont
  {Codello}}\ and\ \bibinfo {author} {\bibfnamefont {R.}~\bibnamefont
  {Percacci}},\ }\bibfield  {title} {\bibinfo {title} {{Fixed points of higher
  derivative gravity}},\ }\href {https://doi.org/10.1103/PhysRevLett.97.221301}
  {\bibfield  {journal} {\bibinfo  {journal} {Phys. Rev. Lett.}\ }\textbf
  {\bibinfo {volume} {97}},\ \bibinfo {pages} {221301} (\bibinfo {year}
  {2006})},\ \Eprint {https://arxiv.org/abs/hep-th/0607128}
  {arXiv:hep-th/0607128} \BibitemShut {NoStop}%
\bibitem [{\citenamefont {Percacci}\ and\ \citenamefont
  {Zanusso}(2010)}]{Percacci:2009fh}%
  \BibitemOpen
  \bibfield  {author} {\bibinfo {author} {\bibfnamefont {R.}~\bibnamefont
  {Percacci}}\ and\ \bibinfo {author} {\bibfnamefont {O.}~\bibnamefont
  {Zanusso}},\ }\bibfield  {title} {\bibinfo {title} {{One loop beta functions
  and fixed points in Higher Derivative Sigma Models}},\ }\href
  {https://doi.org/10.1103/PhysRevD.81.065012} {\bibfield  {journal} {\bibinfo
  {journal} {Phys. Rev. D}\ }\textbf {\bibinfo {volume} {81}},\ \bibinfo
  {pages} {065012} (\bibinfo {year} {2010})},\ \Eprint
  {https://arxiv.org/abs/0910.0851} {arXiv:0910.0851 [hep-th]} \BibitemShut
  {NoStop}%
\bibitem [{\citenamefont {Salvio}\ and\ \citenamefont
  {Strumia}(2014)}]{Salvio:2014soa}%
  \BibitemOpen
  \bibfield  {author} {\bibinfo {author} {\bibfnamefont {A.}~\bibnamefont
  {Salvio}}\ and\ \bibinfo {author} {\bibfnamefont {A.}~\bibnamefont
  {Strumia}},\ }\bibfield  {title} {\bibinfo {title} {{Agravity}},\ }\href
  {https://doi.org/10.1007/JHEP06(2014)080} {\bibfield  {journal} {\bibinfo
  {journal} {JHEP}\ }\textbf {\bibinfo {volume} {06}},\ \bibinfo {pages}
  {080}},\ \Eprint {https://arxiv.org/abs/1403.4226} {arXiv:1403.4226 [hep-ph]}
  \BibitemShut {NoStop}%
\bibitem [{\citenamefont {Alvarez-Gaume}\ \emph {et~al.}(2016)\citenamefont
  {Alvarez-Gaume}, \citenamefont {Kehagias}, \citenamefont {Kounnas},
  \citenamefont {L\"ust},\ and\ \citenamefont
  {Riotto}}]{Alvarez-Gaume:2015rwa}%
  \BibitemOpen
  \bibfield  {author} {\bibinfo {author} {\bibfnamefont {L.}~\bibnamefont
  {Alvarez-Gaume}}, \bibinfo {author} {\bibfnamefont {A.}~\bibnamefont
  {Kehagias}}, \bibinfo {author} {\bibfnamefont {C.}~\bibnamefont {Kounnas}},
  \bibinfo {author} {\bibfnamefont {D.}~\bibnamefont {L\"ust}},\ and\ \bibinfo
  {author} {\bibfnamefont {A.}~\bibnamefont {Riotto}},\ }\bibfield  {title}
  {\bibinfo {title} {{Aspects of Quadratic Gravity}},\ }\href
  {https://doi.org/10.1002/prop.201500100} {\bibfield  {journal} {\bibinfo
  {journal} {Fortsch. Phys.}\ }\textbf {\bibinfo {volume} {64}},\ \bibinfo
  {pages} {176} (\bibinfo {year} {2016})},\ \Eprint
  {https://arxiv.org/abs/1505.07657} {arXiv:1505.07657 [hep-th]} \BibitemShut
  {NoStop}%
\bibitem [{\citenamefont {Anselmi}\ and\ \citenamefont
  {Piva}(2018)}]{Anselmi:2018ibi}%
  \BibitemOpen
  \bibfield  {author} {\bibinfo {author} {\bibfnamefont {D.}~\bibnamefont
  {Anselmi}}\ and\ \bibinfo {author} {\bibfnamefont {M.}~\bibnamefont {Piva}},\
  }\bibfield  {title} {\bibinfo {title} {{The Ultraviolet Behavior of Quantum
  Gravity}},\ }\href {https://doi.org/10.1007/JHEP05(2018)027} {\bibfield
  {journal} {\bibinfo  {journal} {JHEP}\ }\textbf {\bibinfo {volume} {05}},\
  \bibinfo {pages} {027}},\ \Eprint {https://arxiv.org/abs/1803.07777}
  {arXiv:1803.07777 [hep-th]} \BibitemShut {NoStop}%
\bibitem [{\citenamefont {Rachwal}\ \emph {et~al.}(2021)\citenamefont
  {Rachwal}, \citenamefont {Modesto}, \citenamefont {Pinzul},\ and\
  \citenamefont {Shapiro}}]{Rachwal:2021bgb}%
  \BibitemOpen
  \bibfield  {author} {\bibinfo {author} {\bibfnamefont {L.}~\bibnamefont
  {Rachwal}}, \bibinfo {author} {\bibfnamefont {L.}~\bibnamefont {Modesto}},
  \bibinfo {author} {\bibfnamefont {A.}~\bibnamefont {Pinzul}},\ and\ \bibinfo
  {author} {\bibfnamefont {I.~L.}\ \bibnamefont {Shapiro}},\ }\bibfield
  {title} {\bibinfo {title} {{Renormalization group in six-derivative quantum
  gravity}},\ }\href {https://doi.org/10.1103/PhysRevD.104.085018} {\bibfield
  {journal} {\bibinfo  {journal} {Phys. Rev. D}\ }\textbf {\bibinfo {volume}
  {104}},\ \bibinfo {pages} {085018} (\bibinfo {year} {2021})},\ \Eprint
  {https://arxiv.org/abs/2104.13980} {arXiv:2104.13980 [hep-th]} \BibitemShut
  {NoStop}%
\bibitem [{\citenamefont {Tseytlin}(2023)}]{Tseytlin:2022flu}%
  \BibitemOpen
  \bibfield  {author} {\bibinfo {author} {\bibfnamefont {A.~A.}\ \bibnamefont
  {Tseytlin}},\ }\bibfield  {title} {\bibinfo {title} {{Comments on a
  4-derivative scalar theory in 4 dimensions}},\ }\href
  {https://doi.org/10.1134/S0040577923120139} {\bibfield  {journal} {\bibinfo
  {journal} {Theor. Math. Phys.}\ }\textbf {\bibinfo {volume} {217}},\ \bibinfo
  {pages} {1969} (\bibinfo {year} {2023})},\ \Eprint
  {https://arxiv.org/abs/2212.10599} {arXiv:2212.10599 [hep-th]} \BibitemShut
  {NoStop}%
\bibitem [{\citenamefont {Grasso}\ and\ \citenamefont
  {Kuzenko}(2023)}]{GrassoKuzenko2023}%
  \BibitemOpen
  \bibfield  {author} {\bibinfo {author} {\bibfnamefont {D.}~\bibnamefont
  {Grasso}}\ and\ \bibinfo {author} {\bibfnamefont {S.}~\bibnamefont
  {Kuzenko}},\ }\bibfield  {title} {\bibinfo {title} {{Effective actions in
  supersymmetric gauge theories: heat kernels for non-minimal operators}},\
  }\href {https://doi.org/10.1007/JHEP06(2023)120} {\bibfield  {journal}
  {\bibinfo  {journal} {JHEP}\ }\textbf {\bibinfo {volume} {06}}\bibfield
  {number} {\bibinfo  {number} { (120)}},\ }\Eprint
  {https://arxiv.org/abs/2302.00957} {arXiv:2302.00957 [hep-th]} \BibitemShut
  {NoStop}%
\bibitem [{\citenamefont {Donoghue}\ and\ \citenamefont
  {Menezes}(2023)}]{Donoghue:2023yjt}%
  \BibitemOpen
  \bibfield  {author} {\bibinfo {author} {\bibfnamefont {J.~F.}\ \bibnamefont
  {Donoghue}}\ and\ \bibinfo {author} {\bibfnamefont {G.}~\bibnamefont
  {Menezes}},\ }\bibfield  {title} {\bibinfo {title} {{Higher Derivative Sigma
  Models}},\ }\href@noop {} {\  (\bibinfo {year} {2023})},\ \Eprint
  {https://arxiv.org/abs/2308.13704} {arXiv:2308.13704 [hep-th]} \BibitemShut
  {NoStop}%
\bibitem [{\citenamefont {Holdom}(2023)}]{Holdom:2023usn}%
  \BibitemOpen
  \bibfield  {author} {\bibinfo {author} {\bibfnamefont {B.}~\bibnamefont
  {Holdom}},\ }\bibfield  {title} {\bibinfo {title} {{Running couplings and
  unitarity in a 4-derivative scalar field theory}},\ }\href
  {https://doi.org/10.1016/j.physletb.2023.138023} {\bibfield  {journal}
  {\bibinfo  {journal} {Phys. Lett. B}\ }\textbf {\bibinfo {volume} {843}},\
  \bibinfo {pages} {138023} (\bibinfo {year} {2023})},\ \Eprint
  {https://arxiv.org/abs/2303.06723} {arXiv:2303.06723 [hep-th]} \BibitemShut
  {NoStop}%
\bibitem [{\citenamefont {Buccio}\ \emph {et~al.}(2024)\citenamefont {Buccio},
  \citenamefont {Donoghue}, \citenamefont {Menezes},\ and\ \citenamefont
  {Percacci}}]{Buccio:2024hys}%
  \BibitemOpen
  \bibfield  {author} {\bibinfo {author} {\bibfnamefont {D.}~\bibnamefont
  {Buccio}}, \bibinfo {author} {\bibfnamefont {J.~F.}\ \bibnamefont
  {Donoghue}}, \bibinfo {author} {\bibfnamefont {G.}~\bibnamefont {Menezes}},\
  and\ \bibinfo {author} {\bibfnamefont {R.}~\bibnamefont {Percacci}},\
  }\bibfield  {title} {\bibinfo {title} {{Physical Running of Couplings in
  Quadratic Gravity}},\ }\href {https://doi.org/10.1103/PhysRevLett.133.021604}
  {\bibfield  {journal} {\bibinfo  {journal} {Phys. Rev. Lett.}\ }\textbf
  {\bibinfo {volume} {133}},\ \bibinfo {pages} {021604} (\bibinfo {year}
  {2024})},\ \Eprint {https://arxiv.org/abs/2403.02397} {arXiv:2403.02397
  [hep-th]} \BibitemShut {NoStop}%
\bibitem [{\citenamefont {Barvinsky}\ and\ \citenamefont
  {Wachowski}(2022)}]{Wach3}%
  \BibitemOpen
  \bibfield  {author} {\bibinfo {author} {\bibfnamefont {A.~O.}\ \bibnamefont
  {Barvinsky}}\ and\ \bibinfo {author} {\bibfnamefont {W.}~\bibnamefont
  {Wachowski}},\ }\bibfield  {title} {\bibinfo {title} {{Heat kernel expansion
  for higher order minimal and nonminimal operators}},\ }\href
  {https://doi.org/10.1103/PhysRevD.105.065013} {\bibfield  {journal} {\bibinfo
   {journal} {Phys. Rev.}\ }\textbf {\bibinfo {volume} {D105}},\ \bibinfo
  {pages} {065013} (\bibinfo {year} {2022})},\ \Eprint
  {https://arxiv.org/abs/2112.03062} {arXiv:2112.03062 [hep-th]} \BibitemShut
  {NoStop}%
\bibitem [{\citenamefont {Carinhas}\ and\ \citenamefont
  {Fulling}(1990)}]{Fulling}%
  \BibitemOpen
  \bibfield  {author} {\bibinfo {author} {\bibfnamefont {P.~A.}\ \bibnamefont
  {Carinhas}}\ and\ \bibinfo {author} {\bibfnamefont {S.~A.}\ \bibnamefont
  {Fulling}},\ }\bibfield  {title} {\bibinfo {title} {{Computational
  asymptotics of fourth-order operators}},\ }in\ \href@noop {} {\emph {\bibinfo
  {booktitle} {Asymptotic and computational analysis: conference in honor of
  Frank W.J. Olver's 65th birthday}}},\ \bibinfo {series and number} {Proc.
  Sympos. Pure Math.}\ (\bibinfo  {publisher} {MARCEL DEKKER, Inc.},\ \bibinfo
  {address} {New York},\ \bibinfo {year} {1990})\BibitemShut {NoStop}%
\bibitem [{\citenamefont {Gusynin}(1989)}]{Gusynin1989}%
  \BibitemOpen
  \bibfield  {author} {\bibinfo {author} {\bibfnamefont {V.~P.}\ \bibnamefont
  {Gusynin}},\ }\bibfield  {title} {\bibinfo {title} {{New algorithm for
  computing the coefficients in the heat kernel expansion}},\ }\href
  {https://doi.org/10.1016/0370-2693(89)90811-3} {\bibfield  {journal}
  {\bibinfo  {journal} {Phys. Lett.}\ }\textbf {\bibinfo {volume} {B225}},\
  \bibinfo {pages} {233} (\bibinfo {year} {1989})}\BibitemShut {NoStop}%
\bibitem [{\citenamefont {Gusynin}(1990)}]{Gusynin1990}%
  \BibitemOpen
  \bibfield  {author} {\bibinfo {author} {\bibfnamefont {V.~P.}\ \bibnamefont
  {Gusynin}},\ }\bibfield  {title} {\bibinfo {title} {{Seeley--Gilkey
  coefficients for fourth-order operators on a Riemannian manifold}},\ }\href
  {https://doi.org/10.1016/0550-3213(90)90233-4} {\bibfield  {journal}
  {\bibinfo  {journal} {Nucl. Phys.}\ }\textbf {\bibinfo {volume} {B333}},\
  \bibinfo {pages} {296} (\bibinfo {year} {1990})}\BibitemShut {NoStop}%
\bibitem [{\citenamefont {Gusynin}(1991)}]{Gusynin1991}%
  \BibitemOpen
  \bibfield  {author} {\bibinfo {author} {\bibfnamefont {V.~P.}\ \bibnamefont
  {Gusynin}},\ }\bibfield  {title} {\bibinfo {title} {{Asymptotics of the heat
  kernel for nonminimal differential operators}},\ }\href
  {https://doi.org/10.1007/BF01067283} {\bibfield  {journal} {\bibinfo
  {journal} {Ukr. Math. J.}\ }\textbf {\bibinfo {volume} {43}},\ \bibinfo
  {pages} {1432} (\bibinfo {year} {1991})}\BibitemShut {NoStop}%
\bibitem [{\citenamefont {Gusynin}\ and\ \citenamefont
  {Gorbar}(1991)}]{GusyninGorbar}%
  \BibitemOpen
  \bibfield  {author} {\bibinfo {author} {\bibfnamefont {V.~P.}\ \bibnamefont
  {Gusynin}}\ and\ \bibinfo {author} {\bibfnamefont {E.~V.}\ \bibnamefont
  {Gorbar}},\ }\bibfield  {title} {\bibinfo {title} {{Local heat kernel
  asymptotics for nonminimal differential operators}},\ }\href
  {https://doi.org/10.1016/0370-2693(91)91534-3} {\bibfield  {journal}
  {\bibinfo  {journal} {Phys. Lett.}\ }\textbf {\bibinfo {volume} {B270}},\
  \bibinfo {pages} {29} (\bibinfo {year} {1991})}\BibitemShut {NoStop}%
\bibitem [{\citenamefont {Gusynin}\ \emph {et~al.}(1991)\citenamefont
  {Gusynin}, \citenamefont {Gorbar},\ and\ \citenamefont
  {Romankov}}]{GusyninGorbarRomankov}%
  \BibitemOpen
  \bibfield  {author} {\bibinfo {author} {\bibfnamefont {V.~P.}\ \bibnamefont
  {Gusynin}}, \bibinfo {author} {\bibfnamefont {E.~V.}\ \bibnamefont
  {Gorbar}},\ and\ \bibinfo {author} {\bibfnamefont {V.~V.}\ \bibnamefont
  {Romankov}},\ }\bibfield  {title} {\bibinfo {title} {{Heat kernel expansion
  for nonminimal differential operations and manifolds with torsion}},\ }\href
  {https://doi.org/10.1016/0550-3213(91)90568-i} {\bibfield  {journal}
  {\bibinfo  {journal} {Nucl. Phys.}\ }\textbf {\bibinfo {volume} {B362}},\
  \bibinfo {pages} {449} (\bibinfo {year} {1991})}\BibitemShut {NoStop}%
\bibitem [{\citenamefont {Seeley}(1967)}]{Seeley}%
  \BibitemOpen
  \bibfield  {author} {\bibinfo {author} {\bibfnamefont {R.~T.}\ \bibnamefont
  {Seeley}},\ }\bibfield  {title} {\bibinfo {title} {{Complex powers of an
  elliptic operator}},\ }in\ \href {https://doi.org/10.1090/pspum/010} {\emph
  {\bibinfo {booktitle} {Singular Integrals}}},\ \bibinfo {series} {Proc.
  Sympos. Pure Math.}, Vol.~\bibinfo {volume} {10}\ (\bibinfo  {publisher}
  {Amer. Math. Soc.},\ \bibinfo {address} {Chicago, Ill},\ \bibinfo {year}
  {1967})\ pp.\ \bibinfo {pages} {288--307}\BibitemShut {NoStop}%
\bibitem [{\citenamefont {Gilkey}(1975)}]{Gilkey1975}%
  \BibitemOpen
  \bibfield  {author} {\bibinfo {author} {\bibfnamefont {P.~B.}\ \bibnamefont
  {Gilkey}},\ }\bibfield  {title} {\bibinfo {title} {{The spectral geometry of
  a Riemannian manifold}},\ }\href {https://doi.org/10.4310/jdg/1214433164}
  {\bibfield  {journal} {\bibinfo  {journal} {J. Differ. Geom.}\ }\textbf
  {\bibinfo {volume} {10}},\ \bibinfo {pages} {601} (\bibinfo {year}
  {1975})}\BibitemShut {NoStop}%
\bibitem [{\citenamefont {Gilkey}(1979)}]{Gilkey1979}%
  \BibitemOpen
  \bibfield  {author} {\bibinfo {author} {\bibfnamefont {P.~B.}\ \bibnamefont
  {Gilkey}},\ }\bibfield  {title} {\bibinfo {title} {{Recursion relations and
  the asymptotic behavior of the eigenvalues of the Laplacian}},\ }\href@noop
  {} {\bibfield  {journal} {\bibinfo  {journal} {Compositio Math.}\ }\textbf
  {\bibinfo {volume} {38}},\ \bibinfo {pages} {201} (\bibinfo {year}
  {1979})}\BibitemShut {NoStop}%
\bibitem [{\citenamefont {Vassilevich}(2003)}]{Vassil03}%
  \BibitemOpen
  \bibfield  {author} {\bibinfo {author} {\bibfnamefont {D.~V.}\ \bibnamefont
  {Vassilevich}},\ }\bibfield  {title} {\bibinfo {title} {{Heat kernel
  expansion: user's manual}},\ }\href
  {https://doi.org/10.1016/j.physrep.2003.09.002} {\bibfield  {journal}
  {\bibinfo  {journal} {Phys. Rep.}\ }\textbf {\bibinfo {volume} {388}},\
  \bibinfo {pages} {279} (\bibinfo {year} {2003})},\ \Eprint
  {https://arxiv.org/abs/0306138} {arXiv:0306138 [hep-th]} \BibitemShut
  {NoStop}%
\bibitem [{\citenamefont {Fursaev}\ and\ \citenamefont
  {Vassilevich}(2011)}]{Fursaev:2011zz}%
  \BibitemOpen
  \bibfield  {author} {\bibinfo {author} {\bibfnamefont {D.}~\bibnamefont
  {Fursaev}}\ and\ \bibinfo {author} {\bibfnamefont {D.}~\bibnamefont
  {Vassilevich}},\ }\href {https://doi.org/10.1007/978-94-007-0205-9} {\emph
  {\bibinfo {title} {{Operators, Geometry and Quanta}: {Methods of spectral
  geometry in quantum field theory}}}},\ Theoretical and Mathematical Physics\
  (\bibinfo  {publisher} {Springer},\ \bibinfo {address} {Berlin, Germany},\
  \bibinfo {year} {2011})\BibitemShut {NoStop}%
\bibitem [{\citenamefont {Barvinsky}\ \emph {et~al.}(2019)\citenamefont
  {Barvinsky}, \citenamefont {Pronin},\ and\ \citenamefont
  {Wachowski}}]{Wach2}%
  \BibitemOpen
  \bibfield  {author} {\bibinfo {author} {\bibfnamefont {A.~O.}\ \bibnamefont
  {Barvinsky}}, \bibinfo {author} {\bibfnamefont {P.~I.}\ \bibnamefont
  {Pronin}},\ and\ \bibinfo {author} {\bibfnamefont {W.}~\bibnamefont
  {Wachowski}},\ }\bibfield  {title} {\bibinfo {title} {{Heat kernel for
  higher-order differential operators and generalized exponential functions}},\
  }\href {https://doi.org/10.1103/PhysRevD.100.105004} {\bibfield  {journal}
  {\bibinfo  {journal} {Phys. Rev.}\ }\textbf {\bibinfo {volume} {D100}},\
  \bibinfo {pages} {105004} (\bibinfo {year} {2019})},\ \Eprint
  {https://arxiv.org/abs/1908.02161} {arXiv:1908.02161 [hep-th]} \BibitemShut
  {NoStop}%
\bibitem [{\citenamefont {Jack}\ and\ \citenamefont
  {Osborn}(1984)}]{JackOsborn1984}%
  \BibitemOpen
  \bibfield  {author} {\bibinfo {author} {\bibfnamefont {I.}~\bibnamefont
  {Jack}}\ and\ \bibinfo {author} {\bibfnamefont {H.}~\bibnamefont {Osborn}},\
  }\bibfield  {title} {\bibinfo {title} {{Background field calculations in
  curved spacetime (I). General formalism and application to scalar fields}},\
  }\href {https://doi.org/10.1016/0550-3213(84)90067-1} {\bibfield  {journal}
  {\bibinfo  {journal} {Nucl. Phys.}\ }\textbf {\bibinfo {volume} {B234}},\
  \bibinfo {pages} {331} (\bibinfo {year} {1984})}\BibitemShut {NoStop}%
\bibitem [{\citenamefont {Horava}(2009)}]{Horava}%
  \BibitemOpen
  \bibfield  {author} {\bibinfo {author} {\bibfnamefont {P.}~\bibnamefont
  {Horava}},\ }\bibfield  {title} {\bibinfo {title} {{Quantum gravity at a
  {L}ifshitz point}},\ }\href {https://doi.org/10.1103/PhysRevD.79.084008}
  {\bibfield  {journal} {\bibinfo  {journal} {Phys. Rev. D}\ }\textbf {\bibinfo
  {volume} {79}},\ \bibinfo {pages} {084008} (\bibinfo {year} {2009})},\
  \Eprint {https://arxiv.org/abs/0901.3775} {arXiv:0901.3775 [hep-th]}
  \BibitemShut {NoStop}%
\bibitem [{\citenamefont {Barvinsky}\ \emph {et~al.}(2022)\citenamefont
  {Barvinsky}, \citenamefont {Kurov},\ and\ \citenamefont
  {Sibiryakov}}]{BarvinskyKurov2022}%
  \BibitemOpen
  \bibfield  {author} {\bibinfo {author} {\bibfnamefont {A.~O.}\ \bibnamefont
  {Barvinsky}}, \bibinfo {author} {\bibfnamefont {A.~V.}\ \bibnamefont
  {Kurov}},\ and\ \bibinfo {author} {\bibfnamefont {S.~M.}\ \bibnamefont
  {Sibiryakov}},\ }\bibfield  {title} {\bibinfo {title} {{Beta functions of
  $(3+1)$-dimensional projectable Ho\v{r}ava gravity}},\ }\href
  {https://doi.org/10.1103/PhysRevD.105.044009} {\bibfield  {journal} {\bibinfo
   {journal} {Phys. Rev.}\ }\textbf {\bibinfo {volume} {D105}},\ \bibinfo
  {pages} {044009} (\bibinfo {year} {2022})},\ \Eprint
  {https://arxiv.org/abs/2110.14688} {arXiv:2110.14688 [hep-th]} \BibitemShut
  {NoStop}%
\bibitem [{\citenamefont {Martin-Garcia}()}]{xAct}%
  \BibitemOpen
  \bibfield  {author} {\bibinfo {author} {\bibfnamefont {J.~M.}\ \bibnamefont
  {Martin-Garcia}},\ }\href@noop {} {\bibinfo {title} {{xAct: Efficient tensor
  computer algebra for the Wolfram Language}}},\ \bibinfo {howpublished}
  {\url{http://www.xact.es/}}\BibitemShut {NoStop}%
\bibitem [{\citenamefont {Nutma}(2014)}]{Nutma:2013zea}%
  \BibitemOpen
  \bibfield  {author} {\bibinfo {author} {\bibfnamefont {T.}~\bibnamefont
  {Nutma}},\ }\bibfield  {title} {\bibinfo {title} {{xTras : A field-theory
  inspired xAct package for mathematica}},\ }\href
  {https://doi.org/10.1016/j.cpc.2014.02.006} {\bibfield  {journal} {\bibinfo
  {journal} {Comput. Phys. Commun.}\ }\textbf {\bibinfo {volume} {185}},\
  \bibinfo {pages} {1719} (\bibinfo {year} {2014})},\ \Eprint
  {https://arxiv.org/abs/1308.3493} {arXiv:1308.3493 [cs.SC]} \BibitemShut
  {NoStop}%
\end{thebibliography}%

\end{document}